\documentclass[11pt,a4paper]{article} 
\usepackage{jheppub}

\usepackage{amsmath,amssymb,epsfig,bm,amsfonts,yfonts,bbm,mathrsfs}

\relax
\title{Dynamics of entanglement in expanding quantum fields}

\author[a]{J\"{u}rgen Berges,}
\author[a]{Stefan Floerchinger}
\author[b]{and Raju Venugopalan}
\emailAdd{berges@thphys.uni-heidelberg.de}
\emailAdd{floerchinger@thphys.uni-heidelberg.de}
\emailAdd{raju@bnl.gov}
\affiliation[a]{Institut f\"{u}r Theoretische Physik, Universit\"{a}t Heidelberg, Philosophenweg 16, 69120 Heidelberg, Germany}
\affiliation[b]{Physics Department, Brookhaven National Laboratory, Bldg. 510A, Upton, NY 11973, USA}
\abstract{
We develop a novel real-time approach to computing the entanglement between spatial regions for Gaussian states in quantum field theory. The entanglement entropy is characterized in terms of local correlation functions on space-like Cauchy hypersurfaces. The framework is applied to explore an expanding light cone geometry in the particular case of the Schwinger model for quantum electrodynamics in 1+1 space-time dimensions. We observe that the entanglement  entropy becomes extensive in rapidity at early times and that the corresponding local reduced density matrix  is a thermal density matrix for excitations around a coherent field with a time dependent temperature. Since the Schwinger model successfully describes many features of multiparticle production in $e^+ e^-$ collisions, our results provide an attractive explanation in this framework for the apparent thermal nature of multiparticle production even in the absence of significant final state scattering.}
\begin{document}
\maketitle
\section{Introduction}

Entanglement is one of the most puzzling features of quantum theory. It is the subject of intense debates about the foundations of quantum mechanics at least since the influential work of Einstein, Podolsky and Rosen \cite{Einstein:1935rr}. Nowadays entanglement is seen as a key resource for quantum computation and quantum communication devices \cite{NielsenChuang}. Entanglement in quantum field theory was investigated initially mainly with a view on black hole physics \cite{Bombelli:1986rw,Srednicki:1993im,Callan:1994py} and more recently in the context of holography \cite{Ryu:2006bv}. 

For globally pure states, a good measure of the entanglement between a region $A$ and its complement region $B$ is the entanglement entropy. If one considers the reduced density matrix for $A$ that follows from tracing over the Hilbert space associated with region $B$,
\begin{equation}
\rho_A = \text{Tr}_B \rho\,,
\label{eq:reducedDensityMatrix}
\end{equation}
this reduced density matrix is of mixed state form as a result of the entanglement between $A$ and $B$. This can be quantified by the entanglement entropy defined as the von Neumann entropy associated with $\rho_A$,
\begin{equation}
S_A  = - \text{Tr}\{ \rho_A \ln \rho_A \}.
\end{equation}
(More general R\'{e}nyi entanglement entropies will be discussed later in the text.) A detailed understanding of entanglement entropy exists for 1+1 dimensional conformal field theory \cite{Holzhey:1994we,Vidal:2002rm,Korepin:2004zz,Calabrese:2004eu,Calabrese:2009qy}. Technically, a replica trick in the Euclidean formulation of the theory leads to a partition function on an $n$-sheeted Riemann surface which can be evaluated. One can, as a result, compute the entanglement entropy not only of vacuum states but also of finite temperature states \cite{Korepin:2004zz,Calabrese:2004eu}. The formalism has even been extended to discuss nonequilibrium dynamics \cite{Calabrese:2005in,Calabrese:2007mtj,Calabrese:2016xau}. Further, a replica method has also been developed to compute the relative entanglement entropy  between two density matrices \cite{Lashkari:2014yva,Lashkari:2015dia,Ruggiero:2016khg}. 

Novel methods have also been developed for free field theories, in particular for massive (or massless) scalars and fermions \cite{Casini:2005zv,Casini:2007bt,Casini:2009sr}. These methods are based on the Euclidean formulation of these theories and also employ the replica trick. Analytical insights are supplemented by detailed numerical investigations using lattice techniques. In Ref.\ \cite{Casini:2009sr}, alternative real time methods are mentioned but not fully developed. 

In this work, we will be interested in the real-time dynamics of entanglement resulting from the rapid expansion of a system. We will concentrate on Gaussian states and on bosonic theories, a setup that will allow us to discuss the dynamics of an expanding string. The latter, as we shall soon discuss, will be a significant focus of this work. To follow the dynamics in real time, we will use a Schr\"{o}dinger functional formulation for the corresponding density matrix. In order to determine the entanglement entropy of general Gaussian states (equilibrium or nonequilibrium, pure or mixed), it will be convenient to start with a somewhat abstract but fully general discussion of the corresponding mathematics. This discussion follows an approach that goes back to Refs.\ \cite{Bombelli:1986rw, Srednicki:1993im} and develops it further. (See also \cite{Koksma:2010zi,Koksma:2011fx,Sorkin:2012sn, Saravani:2013nwa} for more recent work in this direction.) We will arrive at results that express the von Neumann and R\'{e}nyi entanglement entropies (as well as relative entanglement entropies) directly in terms of traces over combinations of two-point correlation functions within the region $A$. 

Quantum field theory when applied to the temporal evolution of systems, as for example in early universe cosmology or in the presence of time dependent background fields, reveals surprising features \cite{Birrell:1982ix,Mukhanov:2007zz}. For example, excitations can be described with different mode functions; these (and the accompanying creation and annihilation operators corresponding to inequivalent vacuum states) are related by nontrivial Bogoliubov transformations.  In a static Minkowski space problem, one basis is preferred by having mode functions with positive frequencies; this is not the case in time dependent situations with fewer symmetries. It is possible that an incoming vacuum state has nonvanishing particle number with respect to mode functions that are distinguished in having positive frequency states at asymptotically late times. In this case, the time dependence of the background field or of expanding geometry lead effectively to multiparticle production~\cite{Birrell:1982ix,Mukhanov:2007zz,Gelis:2006yv,Gelis:2006cr}.

Entanglement plays an important role in a deeper understanding of such intriguing phenomena. Effective particle production happens typically in terms of entangled Einstein-Podolsky-Rosen pairs. When they separate in space, they contribute to entanglement between spatial regions. This is particularly relevant in the presence of horizons, for example close to a black hole or in the closely related Rindler wedge of spacetime, because causality dictates that observers there cannot recover the full information about a quantum state. As a result, Hawking radiation \cite{Hawking:1974sw} or the closely related Unruh effect \cite{Unruh:1976db} lead to a thermal spectrum of particles. 

While Hawking and Unruh radiation concern idealized static situations governed by an event horizon, similar horizon phenomena can occur in time dependent situations. For example, observations at a given space-time point are, by reasons of causality, only sensitive to the interior of their past light cone. If the quantum field theoretic state is specified on some Cauchy surface in the past of an observer, only regions on this Cauchy surface inside the light cone are of relevance to her. This intersection of the light cone with the Cauchy surface constitutes a kind of particle horizon similar to the cosmological light horizon. 

The interior of this particle horizon typically fills a large volume; in this case, the entanglement entropy, if it scales with the area of its boundary, has a negligible effect. The situation can be very different in an expanding situation and can therefore lead to interesting unanticipated consequences. Further, this phenomenon can help to explain certain puzzling experimental observations in high energy collision experiments that have thus far defied explanation.

A long standing puzzle in elementary electron-positron collisions is that experimental results for particle multiplicities are well described by a thermal model corresponding to Boltzmann weighted distributions with a certain temperature $T$~\cite{Becattini:1995if,Castorina:2007eb,Andronic:2008ev,Becattini:2009fv,Becattini:2010sk}. This is surprising because the theoretical picture we have about these collisions makes thermalization by multiple collisions unlikely. A popular theoretical model for soft QCD processes has been developed in Lund \cite{Andersson:1983ia,Andersson:1998tv} and underlies, for example, the PYTHIA event generator \cite{Sjostrand:2006za,Sjostrand:2007gs}. It is based on expanding QCD strings from which hadrons and resonances are produced by tunneling processes via the Schwinger mechanism. In the standard implementation of the Lund model,  as noted in  ref.~\cite{Fischer:2016zzs}, the thermal-like features seen in experimental data are hard to understand.  One attempt to cure this problem by allowing for fluctuations of the string tension, is discussed in \cite{Bialas:1999zg}. 

\begin{figure}
\begin{center}
\includegraphics[width=0.45\textwidth]{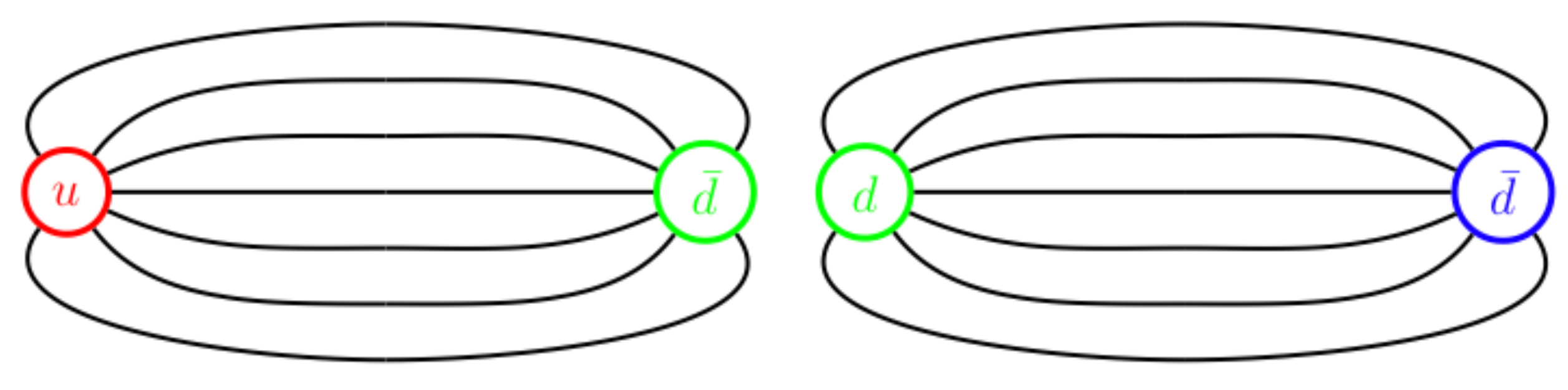}\\
\includegraphics[width=0.45\textwidth]{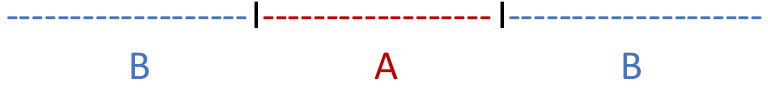}
\caption{Regions in an QCD string.}
\label{figStingPicture}
\end{center}
\end{figure}

In a quantum field theoretic description, different regions in a QCD string are in fact entangled. If one considers a sub-region, for example the region A in fig.\ \ref{figStingPicture}, this interval can be described by a reduced density matrix corresponding to a trace over the complement region B as shown in eq.\ \eqref{eq:reducedDensityMatrix}.
Entanglement between different regions ensures that the reduced density matrix $\rho_A$ is of mixed state form. One may now ask oneself whether this reduced density matrix could resemble locally a thermal state and whether this could explain the close-to-equilibrium distribution of hadron ratios as found experimentally. We will explore this question within the Schwinger model of 1+1-dimensional QED, which is the framework underlying the above mentioned phenomenological Lund model. Remarkably, we find that at very early times, the density matrix for excitations around a coherent field describing the string in a rapidity interval is indeed of thermal form, with a temperature that decays with proper time, $T=\hbar/(2\pi \tau)$ \cite{Berges:2017zws}. This effect is governed by an interplay of entanglement and the expansion dynamics. Our results in this direction are summarized in a recent letter~\cite{Berges:2017zws}. 

We note that in the context of high energy hadron and heavy ion collisions, entanglement and entropy related ideas were discussed some time ago \cite{Elze:1994hj, Elze:1994qa,Muller:2011ra,Kovner:2015hga,Akkelin:2016rhm} as well as more recently \cite{Kharzeev:2017qzs,Martens:2017cvj,Shuryak:2017phz,Muller:2017vnp,Baker:2017wtt}. Besides high energy collider physics, the dynamics of entanglement for quantum field theoretic models is also of relevance in other contexts. One prominent example is of course cosmology \cite{Birrell:1982ix,Mukhanov:2007zz} and  another concerns experiments with ultracold atomic quantum gases where entanglement can actually be investigated directly \cite{FisherInformation,Susceptibilities,MeasuringEntanglementEntropy,EPRsteering}.

This paper is organized as follows. In section \ref{sec:GaussianFieldTheory}, we discuss Gaussian density matrices in quantum field theory and their entanglement properties. We start in section \ref{sec:GaussianPureStates} with a description of pure Gaussian states described at fixed time or on some appropriate Cauchy surface in the Schr\"{o}dinger functional representation. We use a  notation with abstract indices which has the advantage that it can be applied to many concrete physics situations. In section \ref{sec:GaussianDensityMatrices}, we discuss general mixed Gaussian states in terms of a variant of the Glauber-$P$ representation \cite{Glauber:1963tx,Mandel:1995seg} of the density matrix. A key formula here is an expression for the most general Gaussian density matrix in eq.\ \eqref{eq:GeneralGaussianDensityMatrix}. In subsection \ref{sec:ProjectionsAndReducedDensityMatrix}, we discuss the density matrix that emerges when traces over parts of the configuration space are performed. A key result here is that the reduced density matrix remains of Gaussian form with new covariance matrices that can be related to the original covariance matrix by certain projection operators. It is clear that Gaussian density matrices can be fully characterized by expectation values and connected correlation functions of fields and their canonically conjugate momenta and we discuss corresponding relations in subsection \ref{sec:Corrfuncts}. Subsequently, in subsection \ref{sec:Entropies}, we derive expressions for the R\'{e}nyi and von Neumann entropies of general Gaussian density matrices using the replica method at an intermediate step. The resulting expressions are directly governed by connected two-point correlation functions as displayed in eq.\ \eqref{eq:RenyiEntropy} and \eqref{eq:vonNeumannGeneral}. Further simplifications of these expressions can be obtained by the use of canonical transformations, in particular transformations that employ the symplectic structure of fields and their canonical conjugate momenta. We discuss the corresponding mathematics in subsection \ref{sec:Transformations} and emphasize the powerful insight provided by Williamson's theorem into the 
resulting form of the density matrix. A simple representation of the von Neumann entanglement entropy emerges in eq.\ \eqref{eq:EEfromD}.

In section \ref{sec:EntEntropyMinkowski} we illustrate the general formalism of section \ref{sec:GaussianFieldTheory} by applying it to the derivation of the entanglement entropy of an interval of length $L$ in $1+1$ dimensional Minkowski space governed by a free scalar field theory. We consider the corresponding eigenvalue problem in a discrete Fourier representation and discuss the nontrivial role of boundary terms in subsection \ref{sec:eigenvalueproblem}. This is used in subsection \ref{sec:correlatorsfieldmomentum} for the computation of field-field and conjugate momentum correlation functions, and in subsection \ref{sec:entaglemententropyminkoswski} for the computation of the entaglement entropy.

In section~\ref{sec:expandingsystems}, we compute the entanglement entropy and observe that it generates  a locally thermal state for relativistic particles in the rapidly expanding environment of a high energy electron-positron collision. In subsection \ref{sec:SchwingerModel}, we investigate the Schwinger model and in subsections \ref{sec:Generalcoordinatesandbackgroundevolution} and \ref{sec:DynamicsPerturbations} the background field and the dynamics of perturbations of the expanding string solution that capture essential aspects of the collision dynamics. The entanglement entropy of an expanding string is computed in subsection \ref{sec:entaglemententropyexpanding}. We describe the excitations around an expanding coherent field in terms of local density matrices that are of thermal form with a time dependent temperature $T=\hbar / (2\pi \tau)$ in subsection \ref{sec:LocalDensityMatrix} and provide a simple intuitive picture of how the apparent thermal character of these excitations arise. Our conclusions are stated in section \ref{sec:conclusions}. 

Appendix \ref{app:relativeentropy} generalizes our formal discussion of entanglement entropies of Gaussian density matrices to relative entanglement entropies  and appendix \ref{app:A} provides a brief review of the  symmetries, quantum anomalies and bosonization of the Schwinger model for quantum electrodynamics in 1+1 dimensions.

\section{Entropies and entanglement of Gaussian states}
\label{sec:GaussianFieldTheory}

\subsection{Gaussian pure states}
\label{sec:GaussianPureStates}

We shall consider bosonic quantum field operators $\Phi_m$ in the Schr\"odinger picture and employ a set of eigenstates
\begin{equation}
\Phi_m | \phi \rangle = \phi_m | \phi \rangle
\label{eq:eigenstates}
\end{equation}
with eigenvalues $\phi_m$. For a compact notation, the index $m$ can label both continuous degrees of freedom such as position or momentum as well as discrete ones such as spin or internal quantum numbers. In particular, for complex fields, we will also take their complex conjugate to be part of the ``Nambu field'' $\phi_m$.\footnote{For instance, in the absence of any other components, we have $\phi_1 = \phi$ and $\phi_2 = \phi^*$.} Nevertheless, it will often be convenient to also talk about the complex conjugate field $\phi_m^*$ although it is not independent of $\phi_m$. 

If the quantum system is in a pure state $| \psi \rangle$, then $\langle \phi| \psi \rangle$ represents its Schr\"odinger wave functional (to be understood as a functional of the field $\phi$). We will be particularly interested in expanding systems, where at early times the expansion rate can be much larger than typical interaction or scattering rates of field excitations. In this case, the quantum dynamics is typically well described in terms of Gaussian wave functionals, as will be discussed in more detail in section~\ref{sec:SchwingerModel}. The most general form of a Gaussian Schr\"odinger wave functional representing a pure state for a complex bosonic field theory can be written 
as\footnote{For an introductory discussion about most general Gaussian states, see for example~\cite{Berges:2015kfa} (section 2.4) for the case of real bosonic fields.}     
\begin{equation}
\begin{split}
\langle \phi| \psi \rangle \, = \, & \exp\left[ -\frac{1}{2} \phi_m^* h_{mn} \phi_n + \frac{i}{2} \phi_m^* \lambda_m + \frac{i}{2} \lambda^*_m \phi_m +\frac{1}{2} \phi^*_m  \kappa_m + \frac{1}{2} \kappa_m^* \phi_m \right] \\
\, = \, &  \exp\left[ -\frac{1}{2} \phi^\dagger  h \phi + \frac{i}{2} \phi^\dagger \lambda + \frac{i}{2} \lambda^\dagger \phi + \frac{1}{2} \phi^\dagger \kappa + \frac{1}{2} \kappa^\dagger \phi \right].
\end{split}
\label{eq:SWF}
\end{equation}
In the second equation, we have used a condensed notation with $\phi$ denotes the collection of fields, and $h$, $\lambda$ and $\kappa$ are complex quantities that parametrize the state. Their physical meaning will be discussed below. Gaussianity refers to the maximum power of the field appearing in the exponential (\ref{eq:SWF}) being quadratic. 
Correspondingly, we have  
\begin{equation}
\langle \psi | \phi \rangle = \exp\left[ -\frac{1}{2} \phi^\dagger  h^\dagger \phi - \frac{i}{2} \phi^\dagger \lambda - \frac{i}{2} \lambda^\dagger \phi + \frac{1}{2} \phi^\dagger \kappa + \frac{1}{2} \kappa^\dagger \phi \right].
\end{equation}
The norm is given by a functional integral over $\phi$,
\begin{equation}
\begin{split}
\langle \psi | \psi \rangle \, = \, & \int D \phi \exp\left[ -\frac{1}{2} \phi^\dagger (h+h^\dagger) \phi + \phi^\dagger\kappa + \kappa^\dagger\phi \right] \\
\, = \, & \exp\left[ -\frac{1}{2} \ln \det (h+h^\dagger) + 2 \kappa^\dagger (h+h^\dagger)^{-1} \kappa \right] ,
\end{split} 
\end{equation}
where we use $\int D \phi = \prod_n \int D\phi_n$ with $D\phi_n= d \text{Re} \phi_n d \text{Im} \phi_n /\pi$ for complex fields and $ D\phi_n= d \phi_n / \sqrt{\pi}$ for real fields.
Canonical normalization corresponds to this expression being unity. More generally, the scalar product between a state functional $\psi[\phi]$ specified by $h$, $\lambda$ and $\kappa$ and another one $\bar \psi[\phi]$ specified by $\bar h$, $\bar \lambda$ and $\bar \kappa$ is given by the functional integral
\begin{equation}
\langle \bar\psi | \psi \rangle = \int D \phi \; \exp\left[ -\frac{1}{2} \phi^\dagger (h+\bar h^\dagger) \phi + \frac{i}{2} \phi^\dagger (\lambda-\bar \lambda) + \frac{i}{2} (\lambda-\bar \lambda)^\dagger \phi + \frac{1}{2}\phi^\dagger (\kappa+\bar \kappa) + \frac{1}{2} (\kappa+\bar \kappa)^\dagger \phi\right].
\label{eq:scalarProduct}
\end{equation}
Using two sets of eigenstates of (\ref{eq:eigenstates}), whose eigenvalues we denote by $\phi_+$ and $\phi_-$, the pure-state density matrix can be written as
\begin{equation}
\begin{split}
& \rho_{\lambda,\kappa}[\phi_+, \phi_-] \, = \, \langle \phi_+ | \psi \rangle \langle \psi | \phi_- \rangle   \\
& = \, \exp \left[ -\frac{1}{2}  \phi_+^\dagger  h \phi_+    -\frac{1}{2}  \phi_-^\dagger  h^\dagger \phi_- + \frac{i}{2} (\phi_+^\dagger - \phi_-^\dagger) \lambda + \frac{i}{2} \lambda^\dagger (\phi_+ - \phi_-) +\frac{1}{2} (\phi_+^\dagger + \phi_-^\dagger) \kappa + \frac{1}{2}\kappa^\dagger (\phi_+ + \phi_-)  \right].
\end{split}
\label{eq:pureStateDensityMatrix}
\end{equation}
The condition for a pure state density matrix $\text{Tr} \{\rho_{\lambda,\kappa}^2\}/ \text{Tr} \{\rho_{\lambda,\kappa}\}^2=1$ is satisfied as it should be.

We will also work with the canonically conjugate momentum field $\pi_m$. In the Schr\"odinger representation employed, it is given by a functional derivative operator
\begin{equation}
\pi_m = - i \frac{\delta}{\delta \phi_m} \, .
\label{eq:canonicalMomenta}
\end{equation}
The representation \eqref{eq:canonicalMomenta} implies the canonical commutation relations
\begin{equation}
[\phi_m, \pi_m] = i \delta_{mn}, \quad\quad\quad [\phi_m,\phi_n] = [\pi_m,\pi_n]=0,
\end{equation}
and correspondingly for the complex conjugate fields.

\subsection{Gaussian density matrices}
\label{sec:GaussianDensityMatrices}
The Gaussian density matrix we discussed was for a pure state. In the following, we will make the transition to mixed states in terms of the density matrix, but restrict ourselves to situations where the density matrix remains of the Gaussian form\footnote{Deviations from Gaussianity can be treated in this formalism perturbatively but shall not be discussed any further here.}. We  can write such a mixed state density matrix in the form,
\begin{equation}
\rho[\phi_+, \phi_-] = \int D \lambda D\kappa \, P[\lambda,\kappa] \, \rho_{\lambda,\kappa}[\phi_+, \phi_-],
\label{eq:GlauberPDensityMatrix}
\end{equation}
where $\rho_{\lambda,\kappa}[\phi_+, \phi_-]$ is the pure state density matrix in \eqref{eq:pureStateDensityMatrix} (dependent on the fields $\lambda$ and $\kappa$) and $P[\lambda,\kappa]$ is a quasi-probability distribution. When positive, $P[\lambda,\kappa]$ can be seen as the probability distribution for statistical noise in the parameter fields $\lambda$ and $\kappa$.  More generally, however, $P[\lambda,\kappa]$ need not be positive semi-definite. (The density operator $\rho$ should of course be hermitian and positive semi-definite.) Note that eq.\ \eqref{eq:GlauberPDensityMatrix} is closely related (although not identical) to the Glauber-$P$ representation of a density matrix \cite{Glauber:1963tx, Mandel:1995seg}.

In the following, for simplicity, we will take $P[\lambda, \kappa]$ to also be of the Gaussian form,
\begin{equation}
P[\lambda,\kappa] =  \exp\left[ - \begin{pmatrix}\lambda^\dagger - j^\dagger, \kappa^\dagger \end{pmatrix} \Sigma^{-1} \begin{pmatrix}\lambda - j \\ \kappa \end{pmatrix} + \text{const} \right],
\label{eq:GaussianParameterFields}
\end{equation}
with a hermitian operator $\Sigma$ that we take to be of the form\footnote{More general Gaussian forms for $P[\lambda,\kappa]$ are possible but will not be needed for our construction.}
\begin{equation}
\Sigma=\Sigma^\dagger=\begin{pmatrix} \Sigma_a && i \Sigma_b \\ - i \Sigma_b && \Sigma_a \end{pmatrix}, \quad\quad\quad \text{with} \quad \Sigma_a=\Sigma_a^\dagger, \; \Sigma_b= \Sigma_b^\dagger\,.
\end{equation}
For this to be a properly normalizable probability distribution, the eigenvalues of $\Sigma$ should be positive. One may also introduce the linearly transformed parameter fields,
\begin{equation}
\mu_m = \frac{i}{\sqrt{2}} (\lambda_m - j_m) + \frac{1}{\sqrt{2}}\kappa_m, \quad\quad
\nu_m = -\frac{i}{\sqrt{2}} (\lambda_m - j_m)+ \frac{1}{\sqrt{2}}\kappa_m,
\end{equation}
in terms of which the exponent in \eqref{eq:GaussianParameterFields} becomes diagonal,
\begin{equation}
P[\mu,\nu] =  \exp\left[ - \mu^\dagger (\Sigma_a-\Sigma_b)^{-1} \mu - \nu^\dagger (\Sigma_a+\Sigma_b)^{-1} \nu + \text{const} \right].
\label{eq:GaussianParameterFieldsDiagonal}
\end{equation}
Substituting \eqref{eq:GaussianParameterFields} in \eqref{eq:GlauberPDensityMatrix}, one obtains straightforwardly that the mixed state density matrix $\rho[\phi_+, \phi_-]$ is also Gaussian. 
In the limit where $\Sigma \to 0$, the functional $P[\lambda]$ approaches a delta distribution functional and one recovers the pure state we started with. Performing the functional integral over $\lambda$ and an overall shift of fields $\phi_+\to\phi_+-\bar \phi$, $\phi_-\to\phi_- - \bar \phi$ gives 
\begin{equation}
\begin{split}
\rho[\phi_+, \phi_-] = \exp {\Bigg [} & - \frac{1}{2} \begin{pmatrix} \phi_+^\dagger - \bar \phi^\dagger, \phi_-^\dagger - \bar \phi^\dagger \end{pmatrix} \begin{pmatrix} h & - \Sigma_a + \Sigma_b \\ - \Sigma_a - \Sigma_b & h^\dagger \end{pmatrix} \begin{pmatrix} \phi_+ - \bar \phi \\ \phi_- - \bar \phi \end{pmatrix} \\
& + \frac{i}{2} (\phi_+^\dagger - \phi_-^\dagger)j + \frac{i}{2} j^\dagger (\phi_+ - \phi_-) {\Bigg ]}.
\end{split}\label{eq:GeneralGaussianDensityMatrix}
\end{equation}
Interestingly, this is the most general Gaussian density matrix that satisfies the hermiticity property
\begin{equation}
\rho[\phi_+,\phi_-] = \rho[\phi_-,\phi_+]^*.
\label{eq:hermiticityProperty}
\end{equation}
As a consequence, any Gaussian density matrix can be written in the form \eqref{eq:GlauberPDensityMatrix} for suitable Gaussian $P[\lambda,\kappa]$.

The trace of \eqref{eq:GeneralGaussianDensityMatrix} is given by
\begin{equation}
\text{Tr}\{\rho \} = \int D\phi \; \rho[\phi,\phi] = \exp\left[ -\frac{1}{2} \text{ln} \; \text{det}\left(h+h^\dagger - 2\Sigma_a\right) \right] ,
\label{eq:trace}
\end{equation}
and in the following we will assume canonical normalization, $\text{tr}\{ \rho \}=1$, or divide by the appropriate power of the above expression.

\subsection{Projections and reduced density matrix}
\label{sec:ProjectionsAndReducedDensityMatrix}
For the computation of the entanglement entropy and related quantities, we shall require reduced density matrices that result from performing partial traces over some of the degrees of freedom. Towards this end, one may introduce a projection operator $P=P^\dagger$ whereby
\begin{equation}
\tilde \phi_m = (\delta_{mn}-P_{mn}) \phi_n^{+} = (\delta_{mn}-P_{mn}) \phi_n^{-}
\end{equation}
are the fields we want to trace out, and
\begin{equation}
\hat \phi_m^\pm = P_{mn} \phi_n^\pm
\end{equation}
are those we wish to retain. For example, $P$ might be a projector in position space which equals unity in some interval and is zero in the complement region. The reduced density matrix is formally given by
\begin{equation}
\rho_R[\hat \phi_+, \hat \phi_-] = \int D \tilde\phi \; \rho[P \hat \phi^++(1-P)\tilde\phi, P \hat \phi^-+(1-P)\tilde\phi]\,.
\end{equation}
(The projectors have been inserted here for clarity.) For the Gaussian density matrix \eqref{eq:GeneralGaussianDensityMatrix}, one can formally perform the functional integral over $\tilde\phi$ and obtain yet again a Gaussian form for the reduced density matrix,
\begin{equation}
\begin{split}
\rho_R[\hat \phi_+, \hat \phi_-] = \exp {\Bigg [} & - \frac{1}{2} \begin{pmatrix} (\hat\phi_+^\dagger  - \bar \phi^\dagger) P, (\hat\phi_-^\dagger - \bar \phi^\dagger ) P \end{pmatrix} \begin{pmatrix} h -d_{(++)} & - \Sigma_a+\Sigma_b -d_{(+-)} \\ - \Sigma_a - \Sigma_b -d_{(-+)} & h^\dagger - d_{(--)} \end{pmatrix} \begin{pmatrix} P(\phi_+ - \bar \phi) \\ P(\phi_- - \bar \phi) \end{pmatrix} \\
& + \frac{i}{2} (\phi_+^\dagger - \phi_-^\dagger)P j + \frac{i}{2} j^\dagger P (\phi_+ - \phi_-) {\Bigg ]}.
\end{split}\label{eq:ReducedDensityMatrix}
\end{equation}
We have used here the abbreviations
\begin{equation}
\begin{split}
d_{(++)} = &  (h-\Sigma_a+\Sigma_b) (1-P) \left[(1-P)(h+h^\dagger-2\Sigma_a)(1-P)\right]^{-1}(1-P) (h-\Sigma_a-\Sigma_b), \\
d_{(+-)} = &  (h-\Sigma_a+\Sigma_b) (1-P) \left[(1-P)(h+h^\dagger-2\Sigma_a)(1-P)\right]^{-1}(1-P) (h^\dagger -\Sigma_a + \Sigma_b), \\
d_{(-+)} = &  (h^\dagger -\Sigma_a - \Sigma_b) (1-P) \left[(1-P)(h+h^\dagger-2\Sigma_a)(1-P)\right]^{-1}(1-P) (h -\Sigma_a-\Sigma_b) , \\
d_{(--)} = &  (h^\dagger-\Sigma_a-\Sigma_b) (1-P) \left[(1-P)(h+h^\dagger-2\Sigma_a)(1-P)\right]^{-1}(1-P) (h^\dagger -\Sigma_a + \Sigma_b) .
\end{split}\label{eq:apm}
\end{equation}
The operator inversions used here are to be done within the subspace that is integrated out. Note that the reduced density matrix \eqref{eq:ReducedDensityMatrix} is of the same general structure as the original density matrix \eqref{eq:GeneralGaussianDensityMatrix}; in particular, it still satisfies the hermiticity property \eqref{eq:hermiticityProperty}. However if one starts with the density matrix of a pure state where $\Sigma_a=\Sigma_b=0$, the reduced density matrix \eqref{eq:ReducedDensityMatrix} contains in general terms that mix $\phi_+$ and $\phi_-$ (the terms $d_{(+-)}$ and $d_{(-+)}$ in \eqref{eq:apm}) and describes therefore a mixed state, as expected.

Because the reduced density matrix is again Gaussian, it is completely determined by the field expectation values and the two-field correlation functions.  As we will discuss, these can be computed directly on the domain of interest without further reference to projection operators.

\subsection{Correlation functions}
\label{sec:Corrfuncts}

A Gaussian density matrix can be completely characterized in terms of field expectation values and correlation functions of two fields. More specifically, for \eqref{eq:GlauberPDensityMatrix} one has for the field $\phi_m$ and the canonical conjugate momentum field $\pi_m=-i \delta/\delta \phi_m$,
\begin{equation}
\langle \phi_m \rangle = \frac{\int D \phi \, \phi_m \, \rho[\phi, \phi]}{ \int D\phi \, \rho[\phi, \phi]} = \bar \phi_m , \quad\quad\quad \langle \pi_m \rangle =  \frac{\int D \phi \, \left( -i \delta \rho[\phi_+, \phi_-] / \delta \phi_{+m} \right)_{\phi_+=\phi_-=\phi}}{ \int D\phi \, \rho[\phi, \phi]} = j_m^*.
\end{equation}
For the connected correlation functions $\langle A B \rangle_c = \langle AB \rangle - \langle A \rangle \langle B \rangle$ one finds
\begin{equation}
\begin{split}
 \langle \phi_m \phi_n^* \rangle_c =\, & [(h+h^\dagger - 2 \Sigma_a)^{-1}]_{mn}, \\
\langle \pi^*_m \pi_n \rangle_c =\, & [h - (h-\Sigma_a+\Sigma_b) (h+h^\dagger-2\Sigma_a)^{-1} (h-\Sigma_a-\Sigma_b) ]_{mn} \\
= \, & [h^\dagger - (h^\dagger-\Sigma_a-\Sigma_b) (h+h^\dagger - 2 \Sigma_a)^{-1} (h^\dagger - \Sigma_a+\Sigma_b) ]_{mn} \\
= \, & [\Sigma_a - \Sigma_b + (h -\Sigma_a+\Sigma_b) (h+h^\dagger-2\Sigma_a)^{-1} (h^\dagger-\Sigma_a+\Sigma_b)]_{mn} \\
= \, & [\Sigma_a + \Sigma_b + (h^\dagger-\Sigma_a-\Sigma_b)(h+h^\dagger-2\Sigma_a)^{-1} (h-\Sigma_a-\Sigma_b)]_{mn} , \\
\langle \phi_m \pi_n \rangle_c = \, & i [ (h+h^\dagger-2\Sigma_a)^{-1} (h -\Sigma_a - \Sigma_b) ]_{mn}, \\
\langle \pi_n \phi_m \rangle_c = \, & -i [ (h+h^\dagger-2\Sigma_a)^{-1} (h^\dagger -\Sigma_a + \Sigma_b) ]_{mn}, \\
\langle \phi^*_m \pi^*_n \rangle_c = \, & i [ (h -\Sigma_a+\Sigma_b) (h+h^\dagger-2\Sigma_a)^{-1}]_{mn}, \\
\langle \pi^*_n \phi^*_m \rangle_c = \, &- i [ (h^\dagger -\Sigma_a - \Sigma_b) (h+h^\dagger-2\Sigma_a)^{-1}]_{mn}\,.
\end{split}\label{eq:correlations}
\end{equation}
These are compatible with the canonical commutation relations,
\begin{equation}
\begin{split}
& \langle \phi_m \pi_n - \pi_n \phi_m \rangle_c= i \delta_{mn}, \\
& \langle \phi^*_m \pi^*_n - \pi^*_n \phi^*_m \rangle_c= i \delta_{mn}\,.
\end{split}
\end{equation}
Note that the matrices $h$, $h^\dagger$, $\Sigma_a$ and $\Sigma_b$ are fixed in terms of the connected correlation functions of $\phi$, $\phi^*$ and the momenta $\pi$, $\pi^*$. 
For later convenience, we note the relations
\begin{equation}
\begin{split}
a_{mn}\equiv[(h+h^\dagger-2\Sigma_a)^{-1} \Sigma_a]_{mn} = & \sum_k \langle \phi_m \phi_k^* \rangle_c \langle \pi^*_k \pi_n \rangle_c - \frac{1}{4} \delta_{mn}\\
& - \frac{1}{4} \sum_{k,l,r} \langle \phi_m \phi_k^* \rangle_c \langle \phi_k^* \pi_l^* + \pi^*_l \phi_k^*\rangle_c \left[\langle \phi \phi^* \rangle_c \right]^{-1}_{lr} \langle \phi_r \pi_n + \pi_n \phi_r \rangle_c , \\
b_{mn}\equiv[(h+h^\dagger-2\Sigma_a)^{-1} \Sigma_b]_{mn} = \, & \frac{i}{4} \langle \phi_m \pi_n +  \pi_n \phi_m \rangle_c  - \frac{i}{4} \langle \phi_m \phi_k^* \rangle_c \langle \phi_k^* \pi_l^* + \pi_l^* \phi_k^* \rangle_c \left[ \langle \phi \phi^* \rangle_c \right]^{-1}_{ln}.
\end{split}\label{eq:hhdaggerSigma}
\end{equation}
They follow with some algebra from \eqref{eq:correlations}. 

\subsection{Entropy and entanglement entropy}
\label{sec:Entropies}

In this subsection we will determine the R\'{e}nyi entropy for the general Gaussian density matrix \eqref{eq:GeneralGaussianDensityMatrix} and extract from it the von Neumann entropy. The R\'{e}nyi entropy is defined by (we assume standard normalization $\text{tr}\{ \rho \}=1$),
\begin{equation}
S_N(\rho) = \frac{1}{1-N} \ln \text{tr} \{ \rho^N \},
\end{equation}
and the von Neumann entropy follows from this \cite{Holzhey:1994we, Calabrese:2004eu} as a limit
\begin{equation}
S(\rho) = - \text{tr}\{ \rho \;  \text{ln} \; \rho \}= \lim_{N\to 1} S_N(\rho).
\end{equation}
For the purposes of the calculation that follows, it is clear\footnote{This can be seen directly by writing out the expressions for $\text{tr}\{ \rho^N\}$ in the functional integral formalism or more formally by noting that $\bar \phi$ and $j$ can be changed by unitary transformations.} that one can drop $\bar \phi$ and $j$, as they do not enter $\text{tr} \{ \rho^N \}$.  Using otherwise the general expression for $\rho$ in \eqref{eq:GeneralGaussianDensityMatrix} leads after a straight-forward exercise in Gaussian integration to~\cite{Casini:2007bt}
\begin{equation}
\text{Tr} \{ \rho^N \} = \exp \left[ - \frac{1}{2}\text{Tr} \ln \det \left( \mathbbm{M}_N  \right) \right],
\label{eq:trrhoN}
\end{equation}
which contains the $N$ dimensional cyclic matrix (with operator valued entries)
\begin{equation}
\mathbbm{M}_N = (1+2a) \mathbbm{1}_N - (a+b) \mathbbm{Z}_N - (a-b)\mathbbm{Z}_N^T \,.
\end{equation}
We have used here the abbreviations introduced in \eqref{eq:hhdaggerSigma}. We define further $\mathbbm{1}_N$ to be the $N$-dimensional unit matrix and $\mathbbm{Z}_N$ is the $N$-dimensional cyclic matrix $(\mathbbm{Z}_N)_{mn}=\delta_{(m+1)n}$. Here $m,n$ are in the range $1,\ldots, N$ and the index $m=N+1$ is to be identified with the index $m=1$. 

One can write
\begin{equation}
\mathbbm{M}_N = \mathbbm{A}_N(a,b) \,  \mathbbm{A}_N^T (a^T,-b^T)\,,
\end{equation}
with
\begin{equation}
\mathbbm{A}_N(a,b) = \left( \sqrt{\tfrac{1}{4}+a + b^2} + b + \frac{1}{2}  \right) \mathbbm{1}_N - \left( \sqrt{\tfrac{1}{4}+a + b^2} + b - \frac{1}{2}  \right) \mathbbm{Z}_N.
\end{equation}
The determinant of the matrix $\mathbbm{A}_N(a,b)$ is found to be 
\begin{equation}
\text{det} \mathbbm{A}_N(a,b) = \left( \sqrt{\tfrac{1}{4}+a + b^2} + b + \frac{1}{2}  \right)^N - \left( \sqrt{\tfrac{1}{4}+a + b^2} + b - \frac{1}{2}  \right)^N.
\end{equation}

Combining terms leads to a compact expression for the R\'{e}nyi entropy,
\begin{equation}
\begin{split}
S_N(\rho) =  \frac{1}{2(N-1)}{\Bigg \{} & \text{Tr} \ln  \left(  \left( \sqrt{\tfrac{1}{4}+a+b^2}+b+\frac{1}{2} \right)^N - \left( \sqrt{\tfrac{1}{4}+a+b^2}+b-\frac{1}{2} \right)^N \right)  \\
& +\text{Tr} \ln  \left(  \left( \sqrt{\tfrac{1}{4}+a+b^2}-b+\frac{1}{2} \right)^N - \left( \sqrt{\tfrac{1}{4}+a+b^2}-b-\frac{1}{2} \right)^N \right) {\Bigg \} }.
\end{split}
\label{eq:RenyiEntropy}
\end{equation}
We have rederived here in completely general terms, and in the the functional integral representation, a result previously obtained only for a special case in the operator formalism \cite{Holevo1999}.

From the result above, one can directly obtain an expression for the von Neumann entropy,
\begin{equation}
\begin{split}
S = & \frac{1}{2}\text{Tr} \left\{ \left(\sqrt{\frac{1}{4}+a+b^2}+b+\frac{1}{2} \right) \ln \left(\sqrt{\tfrac{1}{4}+a+b^2}+b+\frac{1}{2} \right) \right\} \\
& - \frac{1}{2} \text{Tr} \left\{ \left(\sqrt{\frac{1}{4}+a+b^2}+b-\frac{1}{2}\right) \ln \left(\sqrt{\tfrac{1}{4}+a+b^2}+b-\frac{1}{2}\right) \right\} \\
& + \frac{1}{2}\text{Tr} \left\{ \left(\sqrt{\tfrac{1}{4}+a+b^2}-b+\frac{1}{2} \right) \ln \left(\sqrt{\tfrac{1}{4}+a+b^2}-b+\frac{1}{2} \right) \right\} \\
& - \frac{1}{2} \text{Tr} \left\{ \left(\sqrt{\frac{1}{4}+a+b^2}-b-\frac{1}{2}\right) \ln \left(\sqrt{\tfrac{1}{4}+a+b^2}-b-\frac{1}{2}\right) \right\}\,.
\end{split}
\label{eq:vonNeumannGeneral}
\end{equation}
Note that this is positive semi-definite because $a-b$ and $a+b$ are positive semi-definite. Note also, that $a$ and $b$ can be expressed in terms of correlation functions of fields and canonical momenta using \eqref{eq:hhdaggerSigma}. The above expression simplifies for $b=0$ to
\begin{equation}
\begin{split}
S = & \text{Tr} \left\{ \left(\sqrt{\tfrac{1}{4}+a}+\frac{1}{2} \right) \ln \left(\sqrt{\tfrac{1}{4}+a}+\frac{1}{2} \right) - \left(\sqrt{\tfrac{1}{4}+a}-\frac{1}{2}\right) \ln \left(\sqrt{\tfrac{1}{4}+a}-\frac{1}{2}\right) \right\}\,,
\end{split}
\label{eq:entropyOnlySigmaA}
\end{equation}
and it vanishes as expected for $b=a=0$.

As a first example and check of this formalism, consider a free real massive scalar field in $n$-dimensional  infinite Minkowski space. The correlation functions in this case can be written as
\begin{equation}
\begin{split}
\Delta^S_{\phi\phi}(\vec x-\vec y) = \, & \langle \phi(t,\vec x) \phi(t,\vec y) \rangle =  \int_{\vec p} \frac{e^{i \vec p (\vec x- \vec y)}}{\sqrt{\vec p^2+M^2}} \left[ \frac{1}{2} + n(\vec p) \right], \\
\Delta^S_{\pi\pi}(\vec x-\vec y) = \, & \langle \pi(t,\vec x) \pi(t,\vec y)  \rangle = \int_{\vec p} e^{i \vec p (\vec x- \vec y)} \sqrt{\vec p^2+M^2} \left[ \frac{1}{2} + n(\vec p) \right],
\end{split}\label{eq:DeltaPhiPhiMomentum}
\end{equation}
where $n(\vec p)$ are occupation numbers. The symmetric mixed correlation function vanishes: $\langle \phi \pi + \pi \phi\rangle/2=0$ and therefore $b=0$. Here one can evaluate the (full) entropy easily, because $a=(h+h^\dagger)^{-1} \Sigma_a$ as given in \eqref{eq:hhdaggerSigma} is diagonal in momentum space. One finds
\begin{equation}
S = \int_{\vec p} \left\{ \left(n(\vec p)+1\right) \ln \left(n(\vec p)+1\right) - n(\vec p) \ln \left(n(\vec p)\right)  \right\}\,,
\end{equation}
which is the standard result for free bosonic fields. As expected, the entropy vanishes for $n(\vec p) \to 0$.

We note that similar considerations also hold for relative entropies involving more than one density matrix. This concept is discussed further later and the relative entropy for free scalar fields is discussed in more detail in appendix~\ref{app:relativeentropy}.

\subsection{Symplectic transformations, Williamson's theorem and entanglement entropy}
\label{sec:Transformations}
The above expressions can be further simplified with the help of canonical transformations. One considers unitary changes of the field basis,
\begin{equation}
\begin{split}
\phi_m \to U_{mn} \phi_n\,, & \quad\quad\quad \phi_m^* \to \phi_n^* (U^\dagger)_{nm}\,, \\
\pi_m^* \to  U_{mn} \pi_n^*\,,  & \quad\quad\quad \pi_m \to \pi_n (U^\dagger)_{nm}\,.
\end{split}
\label{eq:UnitaryTransforms}
\end{equation}
They can be used to diagonalize hermitian operators such as $h+h^\dagger$ as is the case in going from position to momentum space. These transformations have unitary representations as transformations of the Schr\"{o}dinger functionals. This is clear as the scalar product \eqref{eq:scalarProduct} remains unchanged by unitary transformations of the field basis due to $D\phi=D(U\phi)$. 

In addition to this, there is a larger class of transformations, which transform fields and momenta into each other. Consider the combined field
\begin{equation}
\chi = \begin{pmatrix} \phi \\ \pi^* \end{pmatrix}, \quad\quad\quad \chi^* = \begin{pmatrix} \phi^* \\ \pi \end{pmatrix}.
\end{equation}
Their canonical commutation relation defines a symplectic metric,
\begin{equation}
[\chi_m, \chi^*_n] =  \Omega_{mn}\,,
\end{equation}
where, symbolically,
\begin{equation}
\Omega = \Omega^\dagger = \begin{pmatrix} 0 && i \mathbbm{1} \\ - i \mathbbm{1} && 0 \end{pmatrix}.
\end{equation}
The transformations
\begin{equation}
\chi_m \to S_{mn} \chi_n, \quad\quad\quad \chi_m^* \to \chi_n^* (S^\dagger)_{nm}\,,
\label{eq:symplecticTransformsFields}
\end{equation}
such that 
\begin{equation}
S \Omega S^\dagger = \Omega ,
\label{eq:conditionSymplecticS}
\end{equation}
are compatible with the canonical commutation relations. This defines a {\it symplectic transformation}. Written in terms of the Lie algebra, $S=\exp[i\theta^A J^A]$, the condition \eqref{eq:conditionSymplecticS} becomes
\begin{equation}
\Omega J^A = (J^A)^\dagger \Omega\,.
\label{eq:specificationSympLieAlgebra}
\end{equation}
Indeed one can confirm that this relation defines a Lie algebra.

Recall that $\phi$ and $\pi$ contain also the corresponding complex conjugate fields so that there are relations
\begin{equation}
\chi_m^* = R_{mn} \chi_n\,, \quad\quad\quad \chi_n = R^{-1}_{nm} \chi_m^* = \chi^*_m (R^\dagger)_{mn}\,.
\end{equation}
One may assume without loss of generality that there is a field basis where all fields are real such that there $R_{mn}=\delta_{mn}$. Of course, the matrix $R$ changes under the unitary, block diagonal transformations \eqref{eq:UnitaryTransforms}. More specifically, one has
\begin{equation}
R \to U^* R (U^\dagger)\,.
\end{equation}
The matrix $R$ transforms also by the symplectic transformations \eqref{eq:symplecticTransformsFields},
\begin{equation}
R \to S^* R (S^{-1})\,.
\end{equation}
Notice that in the field basis where $\phi_m$ is real, there is no change in $R$ for real symplectic transformations. Hence $S_{mp}^*=S_{mp}$, as expected. In a field basis that differs from this by a unitary transformation, the symplectic transformation has different form (and is not necessarily real). Specifically, the symplectic transformation matrix in \eqref{eq:symplecticTransformsFields} and the generator $J^A$ transform under the unitary block diagonal transformations \eqref{eq:UnitaryTransforms} as
\begin{equation}
S \to U S U^\dagger, \quad\quad\quad J^A \to U J^A U^\dagger. 
\end{equation}

We need to show that the symplectic transformations \eqref{eq:symplecticTransformsFields} have unitary representations as transformations of the states of the field theory. This is best done in the field basis where $\phi_m$ are real fields and $R=\mathbbm{1}$. The symplectic transformations are then real and one has
\begin{equation}
J^A = - (J^A)^* = \Omega (J^A)^\dagger \Omega = - \Omega (J^A)^T \Omega\,.
\end{equation}
There is now a representation of the Lie algebra specified by \eqref{eq:specificationSympLieAlgebra} in terms of the operators (\cite{Mukunda:1988cb}, see also \cite{ARL2014})
\begin{equation}
X^A = \frac{1}{2} \chi^*_m \Omega_{mn} (J^A) _{nk} \chi_k\,,
\label{repJA}
\end{equation}
acting on a Schr\"{o}dinger functional. Indeed, one can confirm that they have the same commutation relations as the generators $J^A$. Moreover, one has $(X^A)^\dagger = X^A$ in the sense of the bilinear form \eqref{eq:scalarProduct} so that the symplectic transformations indeed have unitary representations. This is an important result because it allows one to use the symplectic transformations to simplify calculations, for example of the entropy. Because symplectic transformations have unitary representations, they do not change the entropy by construction.

Finally, we note that \eqref{repJA} is invariant under the block diagonal unitary transformations \eqref{eq:UnitaryTransforms} and can therefore be used in any field basis. It is also clear that the corresponding unitary transformation maps Gaussian states to Gaussian states. 

In particular, the hermitian and positive covariance matrix corresponding to the symmetrized correlation function
\begin{equation}
\Delta_{mn} = \frac{1}{2} \left\langle \chi_m \chi_n^* + \chi_n^* \chi_m \right\rangle_c = 
\begin{pmatrix}
\langle \phi_m \phi_n^* \rangle_c &&\tfrac{1}{2}\langle \phi_m \pi_n + \pi_n \phi_m \rangle_c \\
\tfrac{1}{2}\langle \pi_m^* \phi_n^* + \phi_n^* \pi_m^* \rangle_c && \langle \pi_m^* \pi_n \rangle_c \,,
\end{pmatrix},
\label{eq:symCorrFunctChi}
\end{equation}
which is a key ingredient in our discussion of Gaussian states, transforms under symplectic transformations as
\begin{equation}
\Delta \to S \Delta S^\dagger\,.
\end{equation}
This is not a similarity transformation because $S^\dagger \neq S^{-1}$. In other words, the eigenvalues of $\Delta$ are not invariant under symplectic transformations. Williamson's theorem states (see \cite{Simon1999} for a discussion) however that there  must exist a symplectic transformation that brings $\Delta$ to diagonal form,
\begin{equation}
\Delta = \text{diag}(\lambda_1,\lambda_2,\ldots, \lambda_1,\lambda_2,\ldots),
\label{eq:WilliamsonForm}
\end{equation}
with real and positive $\lambda_j>0$. These latter are the {\it symplectic eigenvalues} of the symmetrized covariance matrix.

This is realized by considering the combination $\Delta \Omega$. One can show that it transforms as 
\begin{equation}
\Delta \Omega \to S \Delta S^\dagger \Omega = S \Delta \Omega S^{-1}\,,
\label{eq:transformationDeltaOmegaSymplectic}
\end{equation}
which indeed satisfies the properties of a similarity transformation. The eigenvalues of this combination, $\pm \lambda_j$, are directly related to the symplectic eigenvalues. It is therefore convenient to determine the eigenvalues of $\Delta \Omega$ and to thereby relate observables such as the entanglement entropy to the Williamson form.

We first note that the Williamson form expression \eqref{eq:WilliamsonForm} of the symmetric correlation matrix \eqref{eq:symCorrFunctChi} results in a very simple form for the quantities in \eqref{eq:hhdaggerSigma}:
\begin{equation}
a_{ij} = \left(\lambda_j^2 - \frac{1}{4}\right) \delta_{ij}\,, \quad\quad\quad b_{ij} = 0\,.
\label{eq:sigmaAsigmaBWilliamson}
\end{equation}
Since the Heisenberg uncertainty relation tells us that $a_{ij}$ has to be positive-definite, this indicates that  $\lambda_j\geq 1/2$.

The symplectic transformations we have discussed and the resulting use of Williamson's theorem leads to a very convenient form for entropies. The entropy in \eqref{eq:entropyOnlySigmaA} can be directly expressed in terms of the symplectic eigenvalues as 
\begin{equation}
S = \sum_j \left\{ \left(\lambda_j+\frac{1}{2} \right)\, \ln\left(\lambda_j+\frac{1}{2}\right) - \left(\lambda_j-\frac{1}{2}\right)\, \ln\left(\lambda_j-\frac{1}{2}\right)  \right\}.
\label{eq:W-entropy}
\end{equation}
Moreover, the symplectic eigenvalues $\lambda_m$ follow as pairs of conventional eigenvalues $\pm\lambda_j$ of the combination $\Delta \Omega$. Following Sorkin (\cite{Sorkin:2012sn}, see also \cite{Saravani:2013nwa}), a further simplification can be obtained by considering the matrix 
\begin{equation}
D=\Delta \Omega + \frac{1}{2}\mathbbm{1}\,,
\end{equation}
which has the eigenvalues $\omega_j^+=1/2+\lambda_j$ and $\omega_j^-=1/2-\lambda_j$. As noted, the uncertainty relation gives us $\lambda_j\geq 1/2$; therefore, $\omega_j^+\geq 1$ and $\omega_j^-\leq 0$. One can then write \eqref{eq:W-entropy} simply as 
\begin{equation}
S = \sum_j \left\{ \omega_j^+ \ln(\omega_j^+) + \omega_j^- \ln(-\omega_j^-) \right\}\,,
\end{equation}
where the sum is over pairs of eigenvalues.  More simply,
\begin{equation}
S = \sum_m \omega_m \ln\left(|\omega_m|\right) = \frac{1}{2} \text{Tr}\left\{ D \ln \left(D^2\right) \right\}.
\label{eq:EEfromD}
\end{equation}
In the last expression, the sum is now over all the eigenvalues of $D$. Each negative eigenvalue $\omega_m<0$ is paired with a positive one $1-\omega_m$. A pure state without entropy has $\omega_m\in \{0,1\}$. Finally, we note that the matrix $D$ in symbolic form can be expressed as,
\begin{equation}
D_{mn} = 
\begin{pmatrix}
\tfrac{1}{2} \mathbbm{1} - \frac{i}{2} \langle \phi_m \pi_n + \pi_n \phi_m \rangle_c   &&  i \langle \phi_m \phi_n^* \rangle_c  \\
-i \langle \pi_m^* \pi_n \rangle_c && \tfrac{1}{2}\mathbbm{1}+ \tfrac{i}{2} \langle  \pi_m^* \phi_n^* + \phi_n^* \pi_m^* \rangle_c
\end{pmatrix}=
\begin{pmatrix}
-i \langle \phi_m \pi_n \rangle_c   &&  i \langle \phi_m \phi_n^* \rangle_c  \\
-i \langle \pi_m^* \pi_n \rangle_c &&  i \langle \pi_m^* \phi_n^* \rangle_c
\end{pmatrix}.
\label{eq:DMatrix}
\end{equation}
Hence, the entropy associated with a Gaussian density matrix is fully determined from the set of connected correlation functions of \eqref{eq:DMatrix} evaluated in the domain of interest. It is understood that the operator trace in \eqref{eq:EEfromD} is also restricted to this domain. 

Thus far, we have concentrated on the R\'{e}nyi and von Neumann entropies of a single Gaussian density matrix. It is also possible to determine relative entropies between two Gaussian density matrices $\rho$ and $\sigma$ in a similar way and we discuss this in appendix \ref{app:relativeentropy}. (See ref.\ \cite{Vedral:2002zz} for a general exposition on the concept of relative entropy.) Let us remark here that Williamson's theorem is not as useful for the determination of the relative entropy of two Gaussian density matrices as it is for the determination of the entropy of a single one. This is because it is not guaranteed that there is a basis in which the covariance matrices $\Delta^{(\rho)}$ and $\Delta^{(\sigma)}$ (defined in \eqref{eq:symCorrFunctChi}) simultaneously assume their Williamson diagonal form. However, this should be the case when the matrices $\Delta^{(\rho)} \Omega$ and $\Delta^{(\sigma)} \Omega$ (which transform under symplectic transformations as similarity transforms, see \eqref{eq:transformationDeltaOmegaSymplectic}) commute, i.e.\ $[\Delta^{(\rho)} \Omega, \Delta^{(\sigma)} \Omega] = 0$. One can then write the relative entropy as
\begin{equation}
S(\rho | \sigma) = \sum_m \omega_m^{(\rho)} \left( \ln | \omega_m^{(\sigma)} | - \ln | \omega_m^{(\rho)} | \right),
\end{equation}
where the sum goes over all pairs of simultaneous eigenvalues $(\omega_m^{(\rho)}, \omega_m^{(\sigma)})$ of $D^{(\rho)}=\Delta^{(\rho)} \Omega+\frac{1}{2}\mathbbm{1}$ and $D^{(\sigma)}=\Delta^{(\sigma)} \Omega+\frac{1}{2}\mathbbm{1}$.

\section{Entanglement entropy in Minkowski space}
\label{sec:EntEntropyMinkowski}

In this section, we will illustrate our general formalism by applying it to the derivation of the entanglement entropy of an interval of length $L$ in 1+1 dimensional Minkowski space governed by a free scalar field theory. This is a well studied problem with alternative approaches and results are available from both numerical and analytical methods \cite{Casini:2009sr}. The conventional numerical method to deal with this problem is to discretize the entire theory on a spatial lattice. The interval $L$ corresponds to a finite subset of lattice sites and they are entangled with the lattice sites in the complement region. Numerically, one can study the continuum theory in the limit of finer and finer lattices. The entanglement entropy itself is ultraviolet (UV) divergent and a regulator is provided by the lattice spacing. However universal quantities such as derivatives of the entanglement entropy with respect to the interval length $L$ can be extracted as well. Alternatively, one may consider relative entropies that are finite in the continuum limit as discussed in appendix~\ref{app:relativeentropy}.

The approach we developed in section \ref{sec:GaussianFieldTheory} has the advantage that it depends only on the correlation functions inside the interval considered. Information about entanglement with degrees of freedom outside this interval is taken into account by appropriate boundary conditions in a nontrivial way. While it is rather straightforward to treat a field theory on a finite interval with periodic boundary conditions, which corresponds in fact to an isolated system, it is more involved to properly treat the theory on an interval that is not fully isolated but entangled with a complement region. To illustrate the subtle differences properly is a major focus for the following discussion. 

\subsection{Eigenvalue problem and boundary conditions}
\label{sec:eigenvalueproblem}

We will consider an interval $(-L/2, L/2)$ in Minkowski space with one spatial dimension. A free scalar field $\phi$ will be governed by a Gaussian reduced density matrix on this interval. Moreover, the corresponding matrix entries, namely\ the functions $h$, $h^\dagger$, $\Sigma_a$ and $\Sigma_b$  introduced in section \ref{sec:GaussianFieldTheory}, will be such that the correlation functions have the same form as in infinite space; they are just restricted to the interval. The technical difficulties arise now from the fact that products of these functions involve integrals over the interval $(-L/2,L/2)$, only. For example, the quantity $a=(h+h^\dagger-2\Sigma_a)^{-1}\Sigma_a$ defined in \eqref{eq:hhdaggerSigma} becomes
\begin{equation}
\begin{split}
a(x,y) = & \int_{-L/2}^{L/2} dz \; \Delta^S_{\phi\phi}(x-z)  \Delta^S_{\pi\pi}(z-y) - \frac{1}{4} \delta(x-y) \\
= & \int_{p,q} \left\{ \frac{\sqrt{q^2+M^2}}{4\sqrt{p^2+M^2}}  \frac{\sin\left( \frac{1}{2}(p-q)L \right)}{\frac{1}{2}(p-q)} - \frac{2\pi}{4} \delta(p-q)\right\}  e^{ipx-iqy}.
\end{split}\label{eq:sigmaxy}
\end{equation}
Note that this is a nondiagonal matrix in momentum space for finite $L$. Only in the limit $L\to \infty$ does one obtain $\sin(\frac{1}{2}(p-q) L) / (\frac{1}{2}(p-q)) \to (2\pi)\delta(p-q)$ and $a$ becomes diagonal in momentum space (and zero). The challenge is now to find the eigenvalues of the matrix $a(x,y)$ on the interval $(-L/2,L/2)$.

To solve the eigenvalue problem, we will use a discrete basis involving Fourier expansion on the interval $(-L/2,L/2)$. In doing so, we will not assume periodic boundary conditions. We first divide the relevant function (or field) into a symmetric and an anti-symmetric part,
\begin{equation}
\varphi(x) = \varphi^{(s)}(x) + \varphi^{(a)}(x)\,, \quad\quad \varphi^{(s)}(x)=\frac{\varphi(x)+\varphi(-x)}{2}\,, \quad\quad \varphi^{(a)}(x)=\frac{\varphi(x)-\varphi(-x)}{2}\,.
\end{equation}
The symmetric part can be expanded into a Fourier series
\begin{equation}
\varphi^{(s)}(x) = \frac{1}{L}\sum_{\substack{n=-\infty \\ n \; \text{even}}}^\infty \varphi_n \; e^{in\pi x/L}, \quad\quad\quad \varphi_n = \int_{-L/2}^{L/2} dx \; \varphi^{(s)}(x) \; e^{-in\pi x/L} \quad (n \;  \text{even})\,.
\end{equation}
In a similar fashion, one can expand the anti-symmetric part
\begin{equation}
\varphi^{(a)}(x) = \frac{1}{L}\sum_{\substack{n=-\infty \\ n \; \text{odd}}}^\infty \varphi_n \; e^{in\pi x/L}, \quad\quad\quad \varphi_n = \int_{-L/2}^{L/2} dx \; \varphi^{(a)}(x) \; e^{-in\pi x/L} \quad (n \;  \text{odd}).
\end{equation}
We can summarize this as
\begin{equation}
\varphi(x) = \frac{1}{L}\sum_{n=-\infty}^\infty \varphi_n \; e^{in\pi x/L}, \quad\quad\quad \varphi_n = \int_{-L/2}^{L/2} dx \; \varphi(x) \; \frac{1}{2} \left[e^{-in\pi x/L} + (-1)^n e^{in\pi x/L}\right]\,.
\label{eq:FourierExpansionNoBoundary}
\end{equation}
For $\varphi(x)\in \mathbbm{R}$, one has $\varphi_n= \varphi_{-n}^*$. Note that this type of Fourier expansion does not assume periodic boundary conditions for $\varphi(x)$.

In the limit of large interval length $L\to \infty$, eq. \eqref{eq:FourierExpansionNoBoundary} becomes formally (with $p=n\pi/L$ and $\hat \varphi(q) = \varphi_n$),
\begin{equation}
\varphi(x) = 2 \int \frac{dp}{2\pi} e^{ipx} \hat \varphi(p)\,, \quad\quad\quad \hat \varphi(p) = \int dx \, e^{-ipx}\, \frac{1}{2}\left[\varphi(x) + (-1)^\frac{pL}{\pi} \varphi(-x) \right] \, .
\end{equation}
It is useful to relate these expressions to the standard momentum space representation defined as usual by 
\begin{equation}
\varphi(x) = \int \frac{dq}{2\pi} e^{ipx} \tilde\varphi(p), \quad\quad\quad \tilde\varphi(p) = \int dx \, e^{-ipx} \varphi(x).
\end{equation}
One has for finite interval length $L$
\begin{equation}
\varphi_n = \int \frac{dp}{2\pi} \; \sin(\tfrac{pL}{2}-\tfrac{n\pi}{2}) \left[ \frac{1}{p-\frac{n\pi}{L}} + \frac{1}{p+\frac{n\pi}{L}} \right] \tilde\varphi(p),
\label{eq:phinfromphip}
\end{equation}
and for very large $L$ formally
\begin{equation}
\hat \varphi(p) =\frac{1}{2} \left[ \tilde \varphi(p) + (-1)^\frac{pL}{\pi} \tilde\varphi(-p) \right].
\end{equation}
One recovers the standard Fourier transformed field $2\hat\varphi(p)\to \tilde\varphi(p)$ only if the second, strongly alternating, term above is neglected.

\subsection{Field-field and conjugate momentum correlation functions}
\label{sec:correlatorsfieldmomentum}

We shall now determine the matrix expressions for the correlation functions $\Delta^S_{\phi\phi}(x-y) = \langle \phi(x) \phi(y)\rangle$ and $\Delta^S_{\pi\pi}(x-y)= \langle \pi(x) \pi(y)\rangle$. We first note that the correlation functions in position space can be decomposed as
\begin{equation}
\Delta_{\phi\phi}^S(x-y) = \Delta_{\phi\phi}^{(ss)}(x,y) + \Delta_{\phi\phi}^{(aa)}(x,y), \quad\quad\quad
\Delta_{\pi\pi}^S(x-y) = \Delta_{\pi\pi}^{(ss)}(x,y) + \Delta_{\pi\pi}^{(aa)}(x,y)\,, 
\end{equation}
with
\begin{equation}
\begin{split}
\Delta_{\phi\phi}^{(ss)}(x,y) = & \langle \phi^{(s)}(x) \phi^{(s)}(y)\rangle = \langle \tfrac{1}{2}[\phi(x) + \phi(-x) ] \tfrac{1}{2}[ \phi(y)+ \phi(-y)] \rangle\,, \\
\Delta_{\phi\phi}^{(aa)}(x,y) = & \langle \phi^{(a)}(x) \phi^{(a)}(y)\rangle = \langle \tfrac{1}{2}[\phi(x) - \phi(-x) ] \tfrac{1}{2}[ \phi(y) - \phi(-y)] \rangle\,,
\end{split}
\end{equation}
and similar for the canonical momentum field correlator. The cross terms like $\Delta_{\phi\phi}^{(sa)}$ vanish by parity symmetry. Further, $\Delta_{\phi\phi}^{(ss)}(x,y)$ is symmetric with respect to $x\to -x$ as well as $y \to - y$ while $\Delta_{\phi\phi}^{(aa)}(x,y)$ is anti-symmetric with respect to parity. 

For the combined operator in \eqref{eq:sigmaxy}, it is natural to decompose $a(x,y)=a^{(ss)}(x,y) + a^{(aa)}(x,y)$ with
\begin{equation}
\begin{split}
a^{(ss)}(x,y) = & \int_{-L/2}^{L/2} dz \; \Delta_{\phi\phi}^{(ss)}(x,z) \Delta_{\pi\pi}^{(ss)}(z,y) - \frac{1}{8} \delta(x-y) - \frac{1}{8} \delta(x+y)\,,\\
a^{(aa)}(x,y) = & \int_{-L/2}^{L/2} dz \; \Delta_{\phi\phi}^{(aa)}(x,z) \Delta_{\pi\pi}^{(aa)}(z,y) - \frac{1}{8} \delta(x-y) + \frac{1}{8} \delta(x+y)\,.
\end{split}
\end{equation}
The products of symmetric and anti-symmetric operators vanish under parity symmetry. With respect to the discrete indices $m$ and $n$, one infers that $a_{mn}$ is only nonzero when both indices are even or when both are odd, but that there can be no cross-terms.  

For the momentum space representation \eqref{eq:DeltaPhiPhiMomentum}, using \eqref{eq:phinfromphip} (for $n(\vec p)=0$) one finds that 
\begin{equation}
\begin{split}
\frac{1}{L}[\Delta^{S}_{\phi\phi}]_{m(-n)} = & \frac{1}{L}\int \frac{dp}{2\pi} \; 
\sin(\tfrac{pL}{2}-\tfrac{m\pi}{2}) \sin(\tfrac{pL}{2}-\tfrac{n\pi}{2})
\left[ \frac{1}{p-\frac{m\pi}{L}} + \frac{1}{p+\frac{m\pi}{L}} \right]
\left[ \frac{1}{p-\frac{n\pi}{L}} + \frac{1}{p+\frac{n\pi}{L}} \right] 
\frac{1}{2\sqrt{p^2+M^2}}, \\
\frac{1}{L}[\Delta^S_{\pi\pi}]_{m(-n)} = & \frac{1}{L}\int \frac{dp}{2\pi} \; 
\sin(\tfrac{pL}{2}-\tfrac{m\pi}{2}) \sin(\tfrac{pL}{2}-\tfrac{n\pi}{2})
\left[ \frac{1}{p-\frac{m\pi}{L}} + \frac{1}{p+\frac{m\pi}{L}} \right]
\left[ \frac{1}{p-\frac{n\pi}{L}} + \frac{1}{p+\frac{n\pi}{L}} \right] 
\frac{\sqrt{p^2+M^2}}{2}\,.
\end{split}
\end{equation}
For a state with nonvanishing occupation number, one has to insert factors $(1+2 n(\vec p))$ on the right hand side of these expressions.

For the parity even-even $(ss)$ ($m,n$ even) and odd-odd $(aa)$ ($m,n$ odd) components one has
\begin{equation}
\begin{split}
\frac{1}{L}[\Delta^{(ss)}_{\phi\phi}]_{m(-n)} = & \frac{(-1)^\frac{m-n}{2}}{8\pi L}\int_{-\infty}^\infty dp \; 
\left[ \frac{1}{p-\frac{m\pi}{L}} + \frac{1}{p+\frac{m\pi}{L}} \right]
\left[ \frac{1}{p-\frac{n\pi}{L}} + \frac{1}{p+\frac{n\pi}{L}} \right] 
\frac{1}{\sqrt{p^2+M^2}} \left[1 - \cos(pL)  \right], \\
\frac{1}{L}[\Delta^{(aa)}_{\phi\phi}]_{m(-n)} = & \frac{(-1)^\frac{m-n}{2}}{8\pi L}\int_{-\infty}^\infty dp \; 
\left[ \frac{1}{p-\frac{m\pi}{L}} + \frac{1}{p+\frac{m\pi}{L}} \right]
\left[ \frac{1}{p-\frac{n\pi}{L}} + \frac{1}{p+\frac{n\pi}{L}} \right] 
\frac{1}{\sqrt{p^2+M^2}}\left[1 + \cos(pL)  \right],
\end{split}
\label{eq:phphiCorrelatorIntegral}
\end{equation}
and likewise for $\Delta_{\pi\pi}$, where the square root appears in the numerator. 

In computing these integrals, note first that there are no poles on the real $p$-axis; we can therefore pull the contour slightly below the real axis. In addition, we can write 
\begin{equation}
[1\mp \cos(pL)] = \tfrac{1}{2}\left(1 \mp e^{ipL}\right) + \tfrac{1}{2}\left(1 \mp e^{-ipL}\right)\,.
\end{equation}
The integral involving the first bracket on the r.\ h.\ s.\ can be closed above the real $p$-axis while the second can be closed below. The integral that is closed above picks up a contribution from poles on the real $p$-axis for $m^2=n^2$ as well as a contribution from the branch cut of the square root. The integral that is closed below has contributions from only the branch cut. The contribution from the integral over the poles is simply
\begin{equation}
\frac{\delta_{m-n,0} \pm \delta_{m+n,0}}{4 \sqrt{(\tfrac{m\pi}{L})^2+M^2}}\,.
\label{eq:polcontribution}
\end{equation}
This result is in fact the ground state correlator one would have obtained by quantization of the scalar field theory directly on the interval $(-L/2,L/2)$ with periodic boundary conditions for parity-even modes and anti-periodic boundary conditions for parity-odd modes. It corresponds to a pure state without entanglement. Indeed, keeping only \eqref{eq:polcontribution} together with the corresponding (poles only) approximation for the momentum-momentum correlation function, would lead to a vanishing entanglement entropy. We therefore see concretely from this example that the nontrivial contribution to the correlation function taking entanglement properly into account arises actually from the branch cut contribution to the integral in \eqref{eq:phphiCorrelatorIntegral} and in its counterpart for $\Delta_{\pi\pi}$. 

The contribution from the branch cuts to the field-field correlator, as determined by the right hand side of \eqref{eq:phphiCorrelatorIntegral}, is given by the integral
\begin{equation}
\frac{(-1)^\frac{m-n}{2}L}{\pi} \int_{ML}^\infty dy \frac{y^2}{[y^2+(m\pi)^2][y^2+(n\pi)^2]\sqrt{y^2-(ML)^2}} \left(\pm e^{-y}-1\right)\,.
\end{equation}
We shall first discuss the result without the exponential in the last bracket. This should be a good approximation in particular for $(ML)\gg 1$. For the opposite limit $(ML)\ll 1$, we will add the contribution from the exponential term as well. Performing the integral gives for $m^2\neq n^2$,
\begin{equation}
-\frac{(-1)^\frac{m-n}{2}L}{2\pi^3(m^2-n^2)} \left[ \frac{|m\pi|}{\sqrt{m^2\pi^2 + (ML)^2}} \ln\left( \tfrac{\sqrt{m^2\pi^2+(ML)^2}+|m\pi|}{\sqrt{(m\pi)^2+(ML)^2}-|m\pi|} \right) -
\frac{|n\pi|}{\sqrt{n^2\pi^2 + (ML)^2}} \ln\left( \tfrac{\sqrt{n^2\pi^2+(ML)^2}+|n\pi|}{\sqrt{(n\pi)^2+(ML)^2}-|n\pi|} \right)
 \right],
\end{equation}
while for $m^2=n^2$ the result is
\begin{equation}
-\frac{(-1)^\frac{m-n}{2}L}{4\pi^2 |m|((m\pi)^2+(ML)^2)^{3/2}} \left[ 2\sqrt{m^2\pi^2 (m^2\pi^2 + (ML)^2)} + (ML)^2 \ln \left( \tfrac{\sqrt{m^2\pi^2+(ML)^2}+|m\pi|}{\sqrt{m^2\pi^2+(ML)^2}-|m\pi|} \right) \right].
\end{equation}
Finally, for $m=n=0$ one obtains
\begin{equation}
\frac{(-1)^\frac{m-n}{2}L}{\pi(M L)^2}.
\end{equation}
Adding up these expressions gives the field-field correlation function in the massive case for $ML\gg 1$. 

In the massless case $M=0$, one can also perform all integrals and we obtain the field-field correlators,
\begin{equation}
\begin{split}
\frac{1}{L}[\Delta^{(ss)}_{\phi\phi}]_{m(-n)} = & L \times \begin{cases} 
      \frac{(-1)^\frac{m-n}{2}}{\pi^3} \left[ \frac{\tilde{\text{ci}}(m\pi) - \tilde{\text{ci}}(n\pi) }{m^2-n^2}\right] & \quad (m^2 \neq n^2, m,n \; \text{even}) \\
      \frac{\text{Si}(m\pi)}{2\pi^2 m} & \quad  (m^2=n^2\neq 0, m, n \; \text{even}) \\
   \frac{1}{4\pi} \left[ 3-2 \gamma - 2 \ln(k L) \right] & \quad(m=n=0)\\
      0 & \quad \text{(otherwise)}
    \end{cases} \\
\frac{1}{L}[\Delta^{(aa)}_{\phi\phi}]_{m(-n)} = & L \times \begin{cases}
      \frac{(-1)^\frac{m-n}{2}}{\pi^3} \left[ \frac{\tilde{\text{ci}}(m\pi) - \tilde{\text{ci}}(n\pi) }{m^2-n^2}\right] & \quad (m^2 \neq n^2, m,n \; \text{odd}) \\
      \frac{\text{Si}(m\pi)}{2\pi^2 m} - \frac{1}{\pi^3 m^2}  & \quad  (m=n, m, n \; \text{odd}) \\
      -\frac{\text{Si}(m\pi)}{2\pi^2 m} + \frac{1}{\pi^3 m^2}  & \quad  (m=-n, m, n \; \text{odd}) \\
      0 & \quad \text{(otherwise)}
    \end{cases}
\end{split}
\end{equation}
For the zero mode with $m=n=0$, we have introduced an infrared momentum regulator $k$. Analytic continuation from nonvanishing $m=n$ suggests $k L = e^{\frac{1}{2}-\gamma}$, where $\gamma$ is Euler's constant; we will make this choice in the following. In the above formula, we used the modified cosine integral and sine integral functions given by
\begin{equation}
\begin{split}
\tilde{\text{ci}}(x) = & \text{Ci}(x) - \ln(x) - \gamma =  \int_0^x \frac{\cos(t)-1}{t} =  
\sum_{k=1}^\infty \frac{(-x^2)^k}{2k (2k)!}\,,\\
\text{Si}(x) = & \int_0^x \frac{\sin(t)}{t} .
\end{split}
\end{equation}

Turning now to the canonical momentum field correlator $\Delta_{\pi\pi}$, following steps similar to the above, we find the contribution from poles to be 
\begin{equation}
(\delta_{m-n,0} \pm \delta_{m+n,0}) \frac{1}{4}\sqrt{(\tfrac{m\pi}{L})^2+M^2}\,.
\label{eq:polcontribution2}
\end{equation}
The contribution from the branch cut of the square root gives
\begin{equation}
\frac{(-1)^\frac{m-n}{2}}{\pi L} \int_{ML}^\infty dy \frac{y^2}{[y^2+(m\pi)^2][y^2+(n\pi)^2]} \sqrt{y^2-(ML)^2} \left(1 \mp e^{-y}\right).
\end{equation}
Note that this expression has a logarithmic UV divergence. 

We will follow the same strategy as for the field-field correlation function and  shall discuss first the integral without the exponential term.  This is a good approximation in the massive or long distance limit $(ML)\gg 1$. (Its contribution will be added in the massless limit.) To deal with the UV divergence, we first consider the case $m=n=0$ where one obtains with a UV momentum cutoff $\Lambda$ and assuming $\Lambda \gg M$,
\begin{equation}
\frac{(-1)^\frac{m-n}{2}}{\pi L} \ln\left( \frac{2\Lambda}{e M} \right).
\end{equation}
We subtract now this contribution from the integral and obtain for the remainder when $m^2\neq n^2$,
\begin{equation}
\begin{split}
 \frac{(-1)^\frac{m-n}{2}}{\pi L} & \left( 1 - \frac{1}{2(m^2-n^2)\pi^2}  \left[ |m\pi| \sqrt{m^2 \pi^2 + (ML)^2} \; \ln \left( \tfrac{\sqrt{m^2\pi^2 + (ML)^2}+|m\pi|}{\sqrt{m^2\pi^2+(ML)^2}-|m\pi|} \right) \right.\right. \\
& \left. \left. \quad\quad\quad\quad\quad\quad\quad\quad\quad- |n\pi| \sqrt{n^2 \pi^2 + (ML)^2} \; \ln \left( \tfrac{\sqrt{n^2\pi^2 + (ML)^2}+|n\pi|}{\sqrt{n^2\pi^2+(ML)^2}-|n\pi|} \right)
 \right] \right),
\end{split}
\end{equation}
and for $m^2=n^2$,
\begin{equation}
 \frac{(-1)^\frac{m-n}{2}}{\pi L} \left( \frac{1}{2} - \frac{2m^2 \pi^2 + (ML)^2}{4 |m\pi| \sqrt{m^2 \pi^2 + (ML)^2}} \ln \left( \tfrac{\sqrt{m^2\pi^2 + (ML)^2}+|m\pi|}{\sqrt{m^2\pi^2+(ML)^2}-|m\pi|} \right)  \right).
\end{equation}
By adding these expressions together, one obtains the field-field correlator in the massive case $(ML)\gg 1$. 

One can also perform the integrals for the canonical momentum correlation function in the massless limit to obtain,
\begin{equation}
\begin{split}
\frac{1}{L}[\Delta^{(ss)}_{\pi\pi}]_{m(-n)} = & \frac{1}{L} \times
\begin{cases}
\frac{(-1)^\frac{m-n}{2}}{\pi} \left[ \frac{m^2 \tilde{\text{ci}}(m\pi) - n^2 \tilde{\text{ci}}(n\pi)}{m^2-n^2} + \gamma+ \ln(\Lambda L) \right] & \quad (m^2\neq n^2, m, n \; \text{even})\\
\frac{1}{\pi} \left[ \tilde{\text{ci}}(m\pi) + \tfrac{1}{2} m\pi \, \text{Si}(m\pi) + \gamma + \ln(\Lambda L) \right] & \quad (m^2= n^2 \neq 0, m, n \; \text{even})\\
\frac{1}{\pi} \left[ \gamma + \ln(\Lambda L)  \right] & \quad (m=n=0) \\
0 & \quad (\text{otherwise})
\end{cases} \\
 \frac{1}{L}[\Delta^{(aa)}_{\pi\pi}]_{m(-n)}= & \frac{1}{L} \times
 \begin{cases}
\frac{(-1)^\frac{m+n}{2}}{\pi} \left[ \frac{m^2 \tilde{\text{ci}}(m\pi) - n^2 \tilde{\text{ci}}(n\pi)}{m^2-n^2} + \gamma+ \ln(\Lambda L) \right] & \quad (m^2\neq n^2, m, n \; \text{odd})\\
\frac{1}{\pi} \left[ \tilde{\text{ci}}(m\pi)  + \tfrac{1}{2} m\pi \, \text{Si}(m\pi) -1 +  \gamma + \ln(\Lambda L) \right] & \quad (m= n, m, n \; \text{odd})\\
-\frac{1}{\pi} \left[ \tilde{\text{ci}}(m\pi)  + \tfrac{1}{2} m\pi \, \text{Si}(m\pi) -1 + \gamma + \ln(\Lambda L) \right] & \quad (m= -n, m, n \; \text{odd})\\
0 & \quad (\text{otherwise})
\end{cases}
\end{split}
\end{equation}
Here again $\Lambda$ is the UV momentum cutoff that we introduced to regularize the momentum integral. We assume $\Lambda L\gg m,n$. 

\subsection{Entanglement entropy}
\label{sec:entaglemententropyminkoswski}

\begin{figure}
\centering
\includegraphics[width=0.4\textwidth]{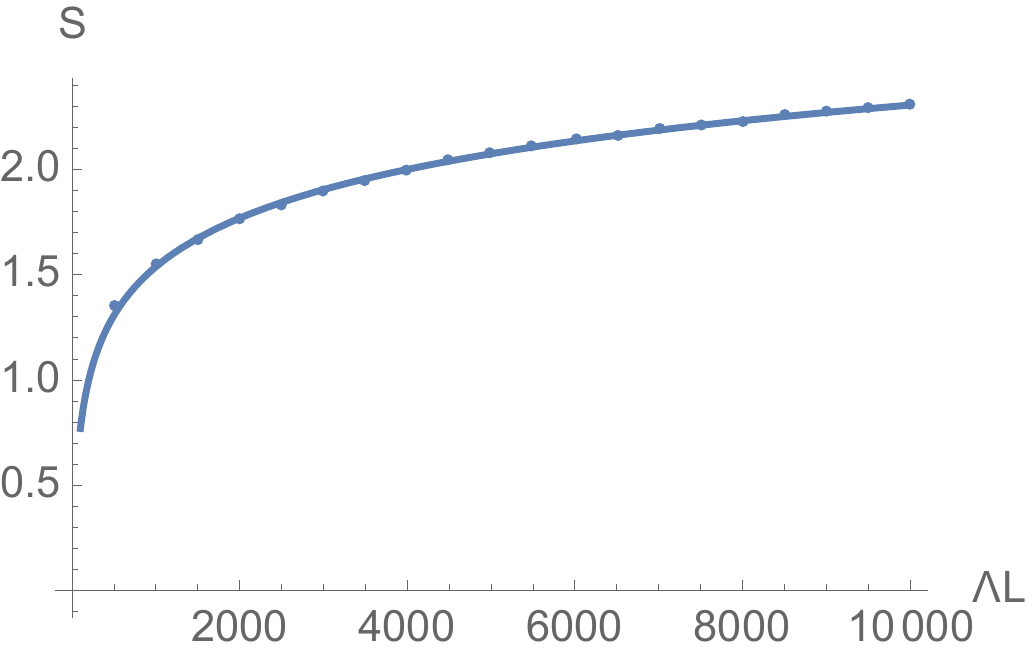}
\caption{Entanglement entropy of a massless bosonic field on a finite interval of length $L$ in Minkowski space vacuum. The points show our numerical result, the curve corresponds to $S=\frac{1}{3}\ln(\Lambda L)+\text{const}$.}
\label{fig1}
\end{figure}

After laying the groundwork above, one may now proceed to determine numerically the eigenvalues of the matrix
\begin{equation}
\frac{1}{L}a_{m(-n)} =  \sum_{l=-\infty}^\infty \left\{ \frac{1}{L}[\Delta^{(ss)}_{\phi\phi}]_{m(-l)} \frac{1}{L}[\Delta^{(ss)}_{\pi\pi}]_{l(-n)} + \frac{1}{L}[\Delta^{(aa)}_{\phi\phi}]_{m(-l)} \frac{1}{L}[\Delta^{(aa)}_{\pi\pi}]_{l(-n)} \right\} - \frac{1}{4} \mathbbm{1}_{m(-n)} ,
\label{eq:sigmaMatrix}
\end{equation}
in an approximation that consists in truncating the infinite sums to a finite range. This can be used to determine the entanglement entropy in terms of \eqref{eq:entropyOnlySigmaA}. Alternatively, one can determine the spectrum of eigenvalues of the matrix $D$ in \eqref{eq:DMatrix} and use \eqref{eq:EEfromD}. Typically $N_\text{max}={\cal O}(100)$ terms are enough. The result for the entanglement entropy in the massless case obtained thus is shown as a function of $\Lambda L$ in Fig.\ \ref{fig1} for $N_\text{max}=300$. We have verified that the result depends only very weakly on the precise value of $N_\text{max}$ as long as $N_\text{max} \ll \Lambda L$. Note that this is the condition under which the expressions for the correlators shown above are valid. Our result shows clearly  that the entanglement entropy for a single massless bosonic field is well described by the anticipated result \cite{Holzhey:1994we,Vidal:2002rm,Korepin:2004zz,Calabrese:2004eu}
\begin{equation}
S=\frac{c}{3} \ln (\Lambda L)+\text{const},
\label{eq:LogLaw}
\end{equation}
with central charge $c=1$, as expected. 

The above formalism can also be used for a massive scalar field although the treatment is numerically more involved. The reason is that now the problem has three length scales given by the interval length $L$, the (inverse of) the UV cutoff $\Lambda^{-1}$ and the inverse of mass $M^{-1}$. An additional UV momentum scale is set by $N_\text{max}/L$. One must be careful to interpolate observables into the physical regime without leaving the range of validity of approximations made in due course. For numerical investigations, it may in this case be advantageous to use a lattice discretization as discussed for example in Ref.~\cite{Casini:2009sr}.
\section{Entanglement entropy for expanding systems}
\label{sec:expandingsystems}

In relativistic dynamics, the separation between a causally connected spacetime region and its acausal complement leads to a natural description of the finite observable region in terms of a reduced density operator. Moreover, the presence of an effective ``lightcone'' for the propagation of signals, the spread of entanglement, and the generation of correlations is also the subject of topical studies in nonrelativistic systems realized with ultracold quantum gases. In these latter cases, the dynamics is often well described in terms of generalized coordinates that reflect the presence of a lightcone. The expressions we will derive in this section will be useful in describing the generation of entanglement entropy in both these cases.

We will compute here the entanglement entropy and obtain its description in terms of a thermal density matrix in the bosonized massless Schwinger model in 1+1-dimensions. As noted previously, this is a popular model whose dynamics is observed to describe key features of particle production in elementary electron-positron collisions. After motivating the massless Schwinger model, we 
we will discuss its bosonized version in terms of massive scalar fields. This will allow us to employ the machinery we developed in the previous sections to compute the time evolution of the entanglement entropy. We will also provide the corresponding results for fermionic fields towards the end of this section.   

\subsection{The Schwinger model}
\label{sec:SchwingerModel}
The Schwinger model describes vector-like QED in $1+1$ space-time dimensions. For a single massive fermion, the Lagrangian is
\begin{equation}
\begin{split}
\mathscr{L} = & - \bar \psi \gamma^\mu (\partial_\mu - i e A_\mu) \psi - m \bar \psi \psi - \tfrac{1}{4}F_{\mu\nu} F^{\mu\nu} \\
 = & \, i \psi_L^* (\partial_0 + \partial_1-ieA_0 - i e A_1) \psi_L + i \psi_R^* (\partial_0 - \partial_1-ie A_0 + ie A_1) \psi_R \\
 & + i m \psi_L^* \psi_R - i m \psi_R^* \psi_L - \tfrac{1}{4}F_{\mu\nu} F^{\mu\nu}. 
\end{split}
\end{equation}
The parameters of the model are the fermion mass $m$ and the U(1) charge $e$. Note that the latter has dimensions of mass in two spacetime dimensions. The parameter $e$ is related to the string tension, as one can see from the following consideration. The energy per unit length of the electric field between charges $e$ and $-e$ is given by $E^2/2$. Moreover, from the Gauss law in 1+1 dimensions one has $E^2=e^2$ such that the string tension is $\sigma=e^2/2$.

As is well known, the Schwinger model in two dimensions can be bosonized exactly \cite{Coleman:1975pw}. (For completeness, we include a discussion of this feature of the model in Appendix \ref{app:A}.) The model can be reformulated as a scalar field theory with Lagrangian (see eq. \eqref{eq:MassiveSchwingerModelBosonizedLagrangian})
\begin{equation}
\mathscr{L} = - \frac{1}{2}\partial_\mu \phi \partial^\mu \phi - \frac{1}{2}\frac{e^2}{\pi} \phi^2 - \frac{m e \exp(\gamma)}{2\pi^{3/2}} \cos\left(2\sqrt{\pi}\phi + \theta\right)\,.
\label{eq:SchwingerModelBosonizedText}
\end{equation} 
Here $\gamma$ is the Euler constant and $\theta$ is the vacuum angle. Note that QED in two dimensions has topologically non-trivial vacuum structure because the gauge group U(1) and the ``boundary of Euclidean space at infinity'' have the same topology $S^1$.

As is clear from the explicit bosonization procedure in appendix \ref{app:A}, the scalar bosons are quadratic in the original field and correspond to fermion-antifermion bound states $\phi\sim \bar \psi\psi$. Hence the fermions themselves are not part of the physical spectrum -- the theory displays (geometric) confinement. 

For the general case of a nonvanishing fermion mass $m$, \eqref{eq:SchwingerModelBosonizedText} is still a nontrivial, interacting theory that cannot be solved easily. In particular, one expects a nontrivial renormalization of the effective potential as well as propagator terms. Also additional bound states that are of quadratic or higher order in $\phi$ could arise. The situation simplifies substantially in the strong interaction limit $e^2\gg m^2$ where one may set $m=0$. The massless Schwinger model becomes a free scalar field theory after bosonization with a scalar boson of mass $M=e/\sqrt{\pi}$. Entanglement in the Schwinger model, as well as in the 't Hooft model, has been investigated for static situations in ref.\ \cite{Goykhman:2015sga}.

\subsection{General coordinates and background evolution}
\label{sec:Generalcoordinatesandbackgroundevolution}

If one considers the free scalar field theory with mass $M$ in arbitrary curved coordinates, one has the action
\begin{equation}
S = \int d^2 x \sqrt{g} \left\{ -\frac{1}{2} g^{\mu\nu}\partial_\mu \phi \partial_\nu \phi - \frac{1}{2} M^2 \phi^2 \right\}.
\end{equation}
For the string that forms between a highly energetic quark-antiquark pair produced in electron-positron collisions, it is natural to consider a boost invariant expansion formulated in Bjorken coordinates with proper time $\tau=\sqrt{(x^0)^2- (x^1)^2}$ and rapidity $\eta = \text{arctanh}(x^1/x^0)$. The metric is $g_{\mu\nu}= \text{diag}(-1,\tau^2)$ and $\sqrt{g}=\tau$. The non-vanishing Christoffel symbols are $\Gamma^\tau_{\eta\eta}=\tau$, $\Gamma^\eta_{\tau\eta}=\Gamma^\eta_{\eta\tau}=\frac{1}{\tau}$.

We will be interested in a situation where the external charges at the endpoints of a string generate a field expectation value $\bar \phi = \langle \phi \rangle$, where the physics dictates that this background field is boost invariant in the rapidity variable. The free-field equation of motion in the Schwinger model for the ``Bjorken expansion" of the $\eta$-independent field expectation value $\bar \phi(\tau)$ is then given by \cite{Loshaj:2011jx}
\begin{equation}
\partial_\tau^2 \bar \phi + \frac{1}{\tau} \partial_\tau \bar \phi + M^2 \bar\phi=0\,.
\label{eq:phiBarEOM}
\end{equation}
The two independent solutions are $\bar\phi(\tau) \sim J_0(M \tau)$ and $\bar \phi(\tau) \sim Y_0(M\tau)$. While the former is regular for $\tau\to 0$, the latter diverges. Both oscillate for late times $\tau\gg 1/M$, which can be understood as multiple string breaking \cite{Hebenstreit:2013baa}. In the limit of vanishing mass $M\to 0$, the corresponding two independent solutions to the equation of motion \eqref{eq:phiBarEOM} are $\bar\phi(\tau) \sim \text{const}$ and $\bar\phi(\tau) \sim \ln(\tau)$. To fulfill the Gauss law, the electric field $E=e\phi/\sqrt{\pi}$ must approach the U(1) charge of the external quarks, $E\to \pm e$ for $\tau \to 0_+$. Therefore we obtain the solutions
\begin{equation}
\bar \phi(\tau) =\pm  \sqrt{\pi} J_0(M\tau).
\end{equation}

The energy-momentum tensor has the form $T^{\mu\nu} = \bar\epsilon u^\mu u^\nu + \bar p (u^\mu u^\nu + g^{\mu\nu})$ with the ``fluid velocity'' $u^\mu=(1,0)$, energy density $\bar \epsilon=\frac{1}{2}(\partial_\tau \bar \phi)^2+\frac{1}{2}M^2 \bar \phi^2$ and ``pressure'' $\bar p=\frac{1}{2}(\partial_\tau \bar \phi)^2-\frac{1}{2}M^2 \bar \phi^2$. One may check easily that the ideal fluid equation of motion,
\begin{equation}
\partial_\tau \bar \epsilon + \frac{\bar \epsilon+ \bar p}{\tau}=0\,,
\end{equation}
is satisfied. However here energy density and pressure are not related by a fixed equation of state as in local thermal equilibrium but they are both determined dynamically by $\bar \phi$. The massless case, $M=0$ is an exception which satisfies the equation of state $\bar p = \bar \epsilon$ for pure radiation in 1+1 dimensions.  More generally, depending on the initial conditions for $\bar \phi$ and $\partial_\tau \bar \phi$, one can have initial conditions between $\bar p=-\bar \epsilon$ and $\bar p=\bar \epsilon$. For the solution in terms of $J_0(M\tau)$, one initially has $\bar p = - \bar \epsilon$ with the result oscillating between this value and $\bar p=\bar \epsilon$. 

Note that the Bjorken metric $g_{\mu\nu}= \text{diag}(-1,\tau^2)$ does not have a Killing vector field pointing in $\tau$ direction; there is no solution $\xi^\mu=(f(\tau),0)$ of the equation
\begin{equation}
\mathscr{L}_\xi g_{\mu\nu} = \xi^\alpha \partial_\alpha g_{\mu\nu} + \partial_\mu \xi^\alpha g_{\alpha\nu} + \partial_\nu \xi^\alpha g_{\mu \alpha} = 0\,.
\end{equation}
However, there is a {\it conformal} Killing vector field of this form which is a solution of 
\begin{equation}
\mathscr{L}_\xi g_{\mu\nu} - \lambda g_{\mu\nu}= \xi^\alpha \partial_\alpha g_{\mu\nu} + \partial_\mu \xi^\alpha g_{\alpha\nu} + \partial_\nu \xi^\alpha g_{\mu \alpha}- \lambda g_{\mu\nu} = 0\,,
\label{eq:conformalKillingEqn}
\end{equation}
given by $\xi^\mu=(c\, \tau,0)$ with $\lambda=2c$ with some constant $c$. These observations allow us to  infer that no local thermal equilibrium state is conceivable which has Bjorken boost symmetry except for the case of a conformal field theory. This insight will play a role in our discussion later on the emergence of a local thermal state and the dynamics generating entanglement entropy.  

\subsection{Dynamics of perturbations}
\label{sec:DynamicsPerturbations}

In the following, we consider perturbations or fluctuations around the background solution $\bar\phi$. We begin by writing the field as
\begin{equation}
\phi(\tau, \eta) = \bar \phi(\tau) +  \varphi(\tau, \eta).
\label{eq:fieldsplitting}
\end{equation}
The fluctuation field has the equation of motion 
\begin{equation}
\partial_\tau^2 \varphi(\tau, \eta) + \frac{1}{\tau} \partial_\tau \varphi(\tau, \eta) + \left(M^2-\frac{1}{\tau^2}\frac{\partial^2}{\partial\eta^2}\right) \varphi(\tau, \eta)=0\,.
\end{equation}
We will now discuss the quantization of the field $\varphi$. Time dependent quantization problems of this type are best solved in terms of convenient mode functions. One writes the quantized field as
\begin{equation}
\varphi(\tau,\eta) = \int \frac{d k}{2\pi} \left\{ a(k) f(\tau, |k|) e^{ik\eta} + a^\dagger(k) \, f^*(\tau, |k|) e^{-ik\eta} \right\},
\end{equation}
where $a(k)$ and $a^\dagger(k)$ are annihilation and creation operators.

The mode functions depend only on the magnitude of the wave vector $|k|$ and are solutions to the differential equation 
\begin{equation}
\partial_\tau^2 f(\tau, k) + \frac{1}{\tau} \partial_\tau f(\tau, k) + \left(M^2+\frac{k^2}{\tau^2}\right) f(\tau, k)=0.
\label{eq:DGEModeFunctions}
\end{equation}
The inner product of these mode functions provides the normalization condition 
\begin{equation}
-i f(\tau, k) \tau \partial_\tau f^*(\tau, k) + i f^*(\tau, k) \tau \partial_\tau f(\tau, k) = 1\,.
\label{eq:normalizationF}
\end{equation}
The canonical conjugate momentum field is 
\begin{equation}
\pi_\phi(\tau,\eta) = \tau \partial_\tau \phi(\tau,\eta)\,,
\end{equation}
and the canonical equal-time commutation relations $[\phi(\tau,\eta), \pi_\phi(\tau,\eta^\prime)] = i \delta(\eta-\eta^\prime)$ imply and are implied by the commutator $[a(k), a^\dagger({k^\prime})] = 2\pi \delta(k-k^\prime)$. 

Note that \eqref{eq:DGEModeFunctions} and \eqref{eq:normalizationF} do not fix the mode functions uniquely. The different possibile functions are related by the Bogoliubov transformations,
\begin{equation}
\begin{split}
\bar f(\tau, k) = & \alpha(k) f(\tau, k) + \beta(k) f^*(\tau, k), \quad\quad\quad\;\, \bar f^*(\tau, k) = \alpha^*(k) f^*(\tau, k) + \beta^*(k) f(\tau, k), \\
f(\tau, k) = & \alpha^*(k) \bar f(\tau, k) - \beta(k) \bar f^*(\tau, k), \quad\quad\quad f^*(\tau, k) = \alpha(k) \bar f^*(\tau, k) - \beta^*(k) \bar f(\tau, k), \\
\end{split}
\end{equation}
with $|\alpha(k)|^2-|\beta(k)|^2=1$. The corresponding creation and annihilation operators are related by
\begin{equation}
\begin{split}
\bar a(k) = & \alpha^*(k)a(k) - \beta^*(k) a^\dagger(k), \quad\quad\quad \bar a^\dagger(k) = - \beta(k) a(k) + \alpha(k) a^\dagger(k),  \\
a(k) = & \alpha(k) \bar a(k) + \beta(k) \bar a^\dagger(k), \quad\quad\quad\quad a^\dagger(k) = \beta^*(k) \bar a(k) + \alpha^*(k) \bar a^\dagger(k) .
\end{split}
\end{equation}
The different sets of mode functions correspond to the vacuum states $|\Omega \rangle$ and $|\bar \Omega \rangle$ respectively. These vacuum states contain no excitations in the sense that they are annihilated by the operators $a(k)|\Omega\rangle = 0$ in the formed case and by $\bar a(k)| \bar \Omega\rangle = 0$, in the latter case. Note however that the vacuum $|\Omega \rangle$ might contain excitations with respect to the operator $\bar a(k)$ and the vacuum $|\bar \Omega \rangle$ with respect to $a(k)$. However, the Bogoliubov transformations that connect the different choices are unitary transformations and therefore do not change entropy. Typically, the vacuum with respect to one set of mode functions corresponds to a squeezed state with respect to other sets of mode functions.

In particular, \eqref{eq:DGEModeFunctions} has the two independent solutions $J_{i k}(M\tau)$ and $Y_{ik}(M\tau)$ or, equivalently, $H^{(2)}_{ik}(M\tau)$ and its complex conjugate $H^{(1)}_{-ik}(M\tau)$. The normalized mode functions corresponding to the latter choice are
\begin{equation}
f(\tau, k) =  \frac{\sqrt{\pi}}{2}e^\frac{k\pi}{2} H^{(2)}_{ik}(M\tau), \quad\quad\quad f^*(\tau, k) =  \frac{\sqrt{\pi}}{2}e^\frac{k\pi}{2} H^{(1)}_{-ik}(M\tau).
\label{eq:HankelModeFunctions}
\end{equation}
The set of mode functions $f(\tau,k)$ in \eqref{eq:HankelModeFunctions} is distinguished by being a superposition of positive frequency modes with respect to time $t$ of standard Minkowski spacetime \cite{Birrell:1982ix}. This means that the standard Minkowski vacuum will be also a vacuum with respect to these mode functions in Bjorken coordinates. In the limit of vanishing mass $M\tau\to 0$, the mode function \eqref{eq:HankelModeFunctions} becomes
\begin{equation}
f(\tau, k) \to e^{-ik \ln (\tau M/2)} \frac{ie^\frac{k\pi}{2}\Gamma(i k)}{2\sqrt{\pi}}  - e^{ik \ln (\tau M/2)} \frac{\sqrt{\pi} e^\frac{k\pi}{2} [\cosh(k \pi)-1]}{2\Gamma(1+i k)} \quad\quad(M\tau\to 0)\,.
\end{equation}
One observes that it contains both positive and negative frequency contributions with respect to the logarithm $\ln(\tau)$ of Bjorken time.

An alternative choice of mode functions is
\begin{equation}
\bar f(\tau, k) = \frac{\sqrt{\pi}}{\sqrt{2\sinh(\pi k)}} J_{-ik}(M\tau)\,, \quad\quad\quad \bar f^*(\tau, k) = \frac{\sqrt{\pi}}{\sqrt{2\sinh(\pi k)}} J_{ik}(M\tau)\,.
\label{eq:BesselModeFunctions}
\end{equation}
In the limit of vanishing mass $M\tau\to 0$, the mode function \eqref{eq:BesselModeFunctions} becomes
\begin{equation}
\bar f(\tau, k) \to e^{-ik \ln (\tau M/2)} \frac{\sqrt{\pi}}{\Gamma(1-i k)\sqrt{2 \sinh{k\pi}}}  = e^{-i k \ln (\tau)-i\theta (k, M)} \frac{1}{\sqrt{2k}} \quad\quad(M\tau\to 0)\,.
\label{eq:fbarMassless}
\end{equation}
In this case, we see that it has only positive frequency contributions with respect to $\ln(\tau)$. The phase in the last equation is given by
\begin{equation}
\theta(k, M) = k \ln (M/2) + \arg(\Gamma(1-i k)) .
\label{eq:phaseTheta}
\end{equation}
Note, in particular, that the factor multiplying $k$ diverges in the formal limit $M\to 0$.

The Bogoliubov coefficients that connect the mode functions \eqref{eq:HankelModeFunctions} and \eqref{eq:BesselModeFunctions} are
\begin{equation}
\alpha(k) = \sqrt{\frac{e^{\pi k}}{2 \sinh(\pi k)}}, \quad\quad\quad \beta(k) = \sqrt{\frac{e^{-\pi k}}{2 \sinh(\pi k)}}\,.
\label{eq:BogoliubovCoefficients}
\end{equation}

The Gaussian density matrix  or the vacuum state for this problem can be specified in terms of field expectation values and the connected correlation functions, 
\begin{equation}
\begin{split}
\langle a^\dagger(k) a(k^\prime) \rangle_c & = n(k) \,2\pi \, \delta(k-k^\prime)\,, \\
\langle a(k) a(k^\prime) \rangle_c & = u(k)  \, 2\pi \,\delta(k+k^\prime)\,, \\
\langle a^\dagger(k) a^\dagger(k^\prime) \rangle_c & = u^*(k) \, 2\pi \, \delta(k+k^\prime)\,.
\end{split}
\end{equation}
For example, the vacuum state with $n(k)=u(k) = u^*(k) = 0$ results in the correlation function
\begin{equation}
\langle  \phi(\tau_1, \eta_1) \phi(\tau_2, \eta_2) \rangle_c 
 = \int \frac{dk}{2\pi}  f(\tau_1, k) f^*(\tau_2, k)  \, e^{ik(\eta_1-\eta_2)} .
\label{eq:statCorrFunctBjorkenMinkVacuum}
\end{equation}
It is instructive to compare \eqref{eq:statCorrFunctBjorkenMinkVacuum} with the corresponding Minkowski space expression
\begin{equation}
\langle \phi(x_1) \phi(x_2) \rangle_c = \frac{1}{2}\int_{-\infty}^\infty \frac{d^2 k}{(2\pi)^2} \; e^{-i k^0(x_1^0-x_2^0) + i k^1(x_1^1-x_2^1)} (2\pi) \, \delta(k^2+M^2).
\label{eq:corrFunctMink}
\end{equation}
Employing the integral representation
\begin{equation}
f(\tau, k) e^{ik \eta} = \frac{i}{2\sqrt{\pi}} \int_{-\infty}^\infty d z \; e^{-i M \cosh(z) x^0 - i M \sinh(z) x^1- i k z} ,
\end{equation}
where $x^0=\tau \cosh(\eta)$ and $x^1=\tau \sinh(\eta)$ are standard Minkowski coordinates, one obtains 
\begin{equation}
\langle \phi(x_1) \phi(x_2) \rangle_c  = \frac{1}{4\pi} \text{Re} \int_{-\infty}^\infty dz \; e^{-i M \cosh(z) (x_1^0-x_2^0) - i M \sinh(z) (x_1^1-x_2^1)} .
\end{equation}
This agrees with \eqref{eq:corrFunctMink} after substituting $k^0=\kappa \cosh(z)$, $k^1=-\kappa \sinh(z)$, thereby confirming that the mode functions in \eqref{eq:HankelModeFunctions} are indeed the ones corresponding to the standard Minkowski space vacuum.

Identifying $\pi(\tau,k) = \tau \partial_\tau \phi^*(\tau, k)$, a complete set of connected correlation functions for the vacuum with $n(k)=u(k) = u^*(k)=0$ in momentum space is given by 
\begin{equation}
\begin{split}
\langle \phi(\tau_1, k) \phi^*(\tau_2, k^\prime) \rangle_c = & 2\pi \delta(k-k^\prime) \; f(\tau_1, k) f^*(\tau_2, k), \\
\langle \pi^*(\tau_1, k) \pi(\tau_2, k^\prime) \rangle_c = & 2\pi \delta(k-k^\prime) \; \dot f(\tau_1, k) \dot f^*(\tau_2, k), \\
\langle \phi(\tau_1, k) \pi(\tau_2, k^\prime) \rangle_c = & 2\pi \delta(k-k^\prime)  \; f(\tau_1, k) \dot f^*(\tau_2, k), \\
\langle \pi^*(\tau_1, k) \phi^*(\tau_2, k^\prime) \rangle_c = & 2\pi \delta(k-k^\prime) \; \dot f(\tau_1, k) f^*(\tau_2, k).
\end{split}
\label{eq:426}
\end{equation}
where we have used the abbreviation $\dot f(\tau, k) = \tau \partial_\tau f(\tau,k)$. 
One can directly verify that with the above correlators, the matrix $D$ in \eqref{eq:DMatrix} has pairs of eigenvalues $\{0,1\}$ such that the entropy associated with the entire expanding string is zero. In fact, one does not need the precise form of $f(\tau,k)$ to show this; the normalization condition in  \eqref{eq:normalizationF} alone is sufficient. 

Now with respect to the alternative set of mode functions, with only positive frequency solutions, the state with $n(k)=u(k) = u^*(k)=0$ has the set of correlation functions,
\begin{equation}
\begin{split}
\langle \bar a^\dagger(k) \bar a(k^\prime) \rangle_c & = \bar n(k) \,2\pi \, \delta(k-k^\prime) =  |\beta(k)|^2 \,2\pi \, \delta(k-k^\prime), \\
\langle \bar a(k) \bar a(k^\prime) \rangle_c & = \bar u(k)  \, 2\pi \,\delta(k+k^\prime) = -\alpha^*(k) \beta^*(k)  \, 2\pi \,\delta(k+k^\prime), \\
\langle \bar a^\dagger(k) \bar a^\dagger(k^\prime) \rangle_c & = \bar u^*(k) \, 2\pi \, \delta(k+k^\prime) = -\alpha(k) \beta(k) \, 2\pi \, \delta(k+k^\prime)\,.
\end{split}
\label{eq:abarcorrelationfunctions}
\end{equation}
In this alternative basis, the correlation functions do not look like those of an empty state but rather of one with occupation number $\bar n(k) = |\beta(k)|^2$. From \eqref{eq:BogoliubovCoefficients} one obtains,
\begin{equation}
\bar n(k)= |\beta(k)|^2 = \frac{1}{e^{2\pi k}-1}\,.
\end{equation}
Recalling that the single particle energy in the expanding situation is $E=\sqrt{k^2/\tau^2+M^2}$, for massless bosons, this distribution appearing in the ``diagonal'' elements of the correlation matrix (\ref{eq:abarcorrelationfunctions})
corresponds to a thermal spectrum with the time dependent temperature 
\begin{equation}
T=1/(2\pi \tau)\,.
\label{eq:Ttau}
\end{equation}
Such a thermal interpretation is not possible for a nonvanishing mass $M$, but the fact that the quasiparticles defined by the mode functions $\bar f(\tau, k)$ have a nonvanishing occupation number remains true. Of course, a thermal state would have vanishing entries for the other correlators appearing in (\ref{eq:abarcorrelationfunctions}). The ``off-diagonal occupation function'' is
\begin{equation}
\bar u(k) = \bar u^*(k)= -\frac{1}{2 \sinh(\pi k)}\,.
\end{equation}

One may use the relations in \eqref{eq:abarcorrelationfunctions} to express the correlation functions in  \eqref{eq:426} in the alternative basis,
\begin{equation}
\begin{split}
\langle \phi(\tau_1, k) \phi^*(\tau_2, k^\prime) \rangle_c = & 2\pi \delta(k-k^\prime) \; \left\{ \bar f(\tau_1, k) \bar f^*(\tau_2, k) [1+\bar n(k)] + \bar f^*(\tau_1, k) \bar f(\tau_2, k) \bar n(k) \right. \\
& \quad\quad\quad\quad\quad\quad \left.+ \bar f(\tau_1, k) \bar f(\tau_2, k) \bar u(k) + \bar f^*(\tau_1, k) \bar f^*(\tau_2, k) \bar u^*(k)   \right\}, \\
\langle \pi^*(\tau_1, k) \pi(\tau_2, k^\prime) \rangle_c = & 2\pi \delta(k-k^\prime) \; \left\{ \dot {\bar f}(\tau_1, k) \dot {\bar f}^*(\tau_2, k) [1+\bar n(k)]  + \dot {\bar f}^*(\tau_1, k) \dot {\bar f}(\tau_2, k) \bar n(k) \right. \\
& \quad\quad\quad\quad\quad\quad \left.+ \dot {\bar f}(\tau_1, k) \dot {\bar f}(\tau_2, k) \bar u(k) + \dot {\bar f}^*(\tau_1, k) \dot {\bar f}^*(\tau_2, k) \bar u^*(k) \right\}, \\
\langle \phi(\tau_1, k) \pi(\tau_2, k^\prime) \rangle_c = & 2\pi \delta(k-k^\prime)  \; \left\{ \bar f(\tau_1, k) \dot{\bar f}^*(\tau_2, k) [1+\bar n(k)] + \bar f^*(\tau_1, k) \dot{\bar f}(\tau_2, k) \bar n(k) \right.\\
& \quad\quad\quad\quad\quad\quad \left. \bar f(\tau_1, k) \dot{\bar f}(\tau_2, k) \bar u(k) + \bar f^*(\tau_1, k) \dot{\bar f}^*(\tau_2, k) \bar u^*(k) \right\}, \\
\langle \pi^*(\tau_1, k) \phi^*(\tau_2, k^\prime) \rangle_c = & 2\pi \delta(k-k^\prime) \; \left\{ \dot{\bar f}(\tau_1, k) \bar f^*(\tau_2, k) [1+\bar n(k)] + \dot{\bar f}^*(\tau_1, k) \bar f(\tau_2, k) \bar n(k) \right. \\
& \quad\quad\quad\quad\quad\quad \left. + \dot{\bar f}(\tau_1, k) \bar f(\tau_2, k) \bar u(k) + \dot{\bar f}^*(\tau_1, k) \bar f^*(\tau_2, k) \bar u^*(k)  \right\}.
\end{split}
\label{eq:431}
\end{equation}
From these relations, one obtains the equal time correlation functions by setting $\tau_1=\tau_2=\tau$. Moreover, in the limit $M\tau \ll 1$, one can use \eqref{eq:fbarMassless} for the mode functions $\bar f(\tau, k)$. This gives the correlators 
\begin{equation}
\begin{split}
\langle \phi(\tau, k) \phi^*(\tau, k^\prime) \rangle_c = & 2\pi \delta(k-k^\prime) \frac{1}{|k|} \left\{ \left[\tfrac{1}{2}+\bar n(k) \right] + \cos\left[ 2 k \ln(\tau) + 2 \theta(k, M) \right] \, \bar u(k)   \right\}, \\
\langle \pi^*(\tau, k) \pi(\tau, k^\prime) \rangle_c = & 2\pi \delta(k-k^\prime) |k| \left\{ \left[\tfrac{1}{2}+\bar n(k) \right] + \cos\left[ 2 k \ln(\tau) + 2 \theta(k, M) \right] \, \bar u(k)   \right\}, \\
\langle \phi(\tau, k) \pi(\tau, k^\prime) \rangle_c = & 2\pi \delta(k-k^\prime) \left\{ \tfrac{i}{2} - \sin\left[ 2 k \ln(\tau) + 2 \theta(k, M) \right] \, \bar u(k)   \right\}, \\
\langle \pi^*(\tau, k) \phi^*(\tau, k^\prime) \rangle_c = & 2\pi \delta(k-k^\prime) \left\{ -\tfrac{i}{2} + \sin\left[ 2 k \ln(\tau) + 2 \theta(k, M) \right] \, \bar u(k)   \right\}.
\end{split}
\label{eq:phiphicorrelatorbarbasis}
\end{equation}
Note that the ``off-diagonal occupation functions'' $\bar u(k)$ that appear here are always multiplied with sine or cosine functions that are strongly oscillating with $k$ in the limit $M\tau \to 0$. An interpretation of this term in position (rapidity) space is obtained by Fourier transforming the correlators above back to position (rapidity) space and examining the structure, for instance, of $\langle \phi(\tau, \eta) \phi(\tau, \eta^\prime) \rangle$. The strongly oscillating terms $\sim \cos[2 k \ln(M\tau/2)]$ correspond then to pronounced structures in the rapidity difference $\eta-\eta^\prime$ of the spatial correlators at separation $|\eta-\eta^\prime| \sim 2 |\ln(M\tau/2)|$. This rapidity separation becomes large in the limit $M\tau \to 0$.

\subsection{Entanglement entropy of an expanding string}
\label{sec:entaglemententropyexpanding}

We shall now investigate the entanglement entropy of an interval with length $\Delta\eta$ in rapidity at some fixed Bjorken time $\tau$. Following the discussion in section \ref{sec:EntEntropyMinkowski} for the static Minkowski space case, one has to use a discrete Fourier expansion scheme for the finite interval $(-\Delta\eta/2, \Delta\eta/2)$. As discussed there, to ensure boundary conditions are not restricted, it is most convenient to split the fields into symmetric and anti-symmetric components. The calculation proceeds by determining correlation functions as in eq.\ \eqref{eq:DMatrix} at fixed Bjorken time $\tau$ and in the finite interval. As previously, one can use eq.\ \eqref{eq:EEfromD} to determine the entanglement entropy. 
\begin{figure}
	\centering
	\includegraphics[width=0.58\textwidth]{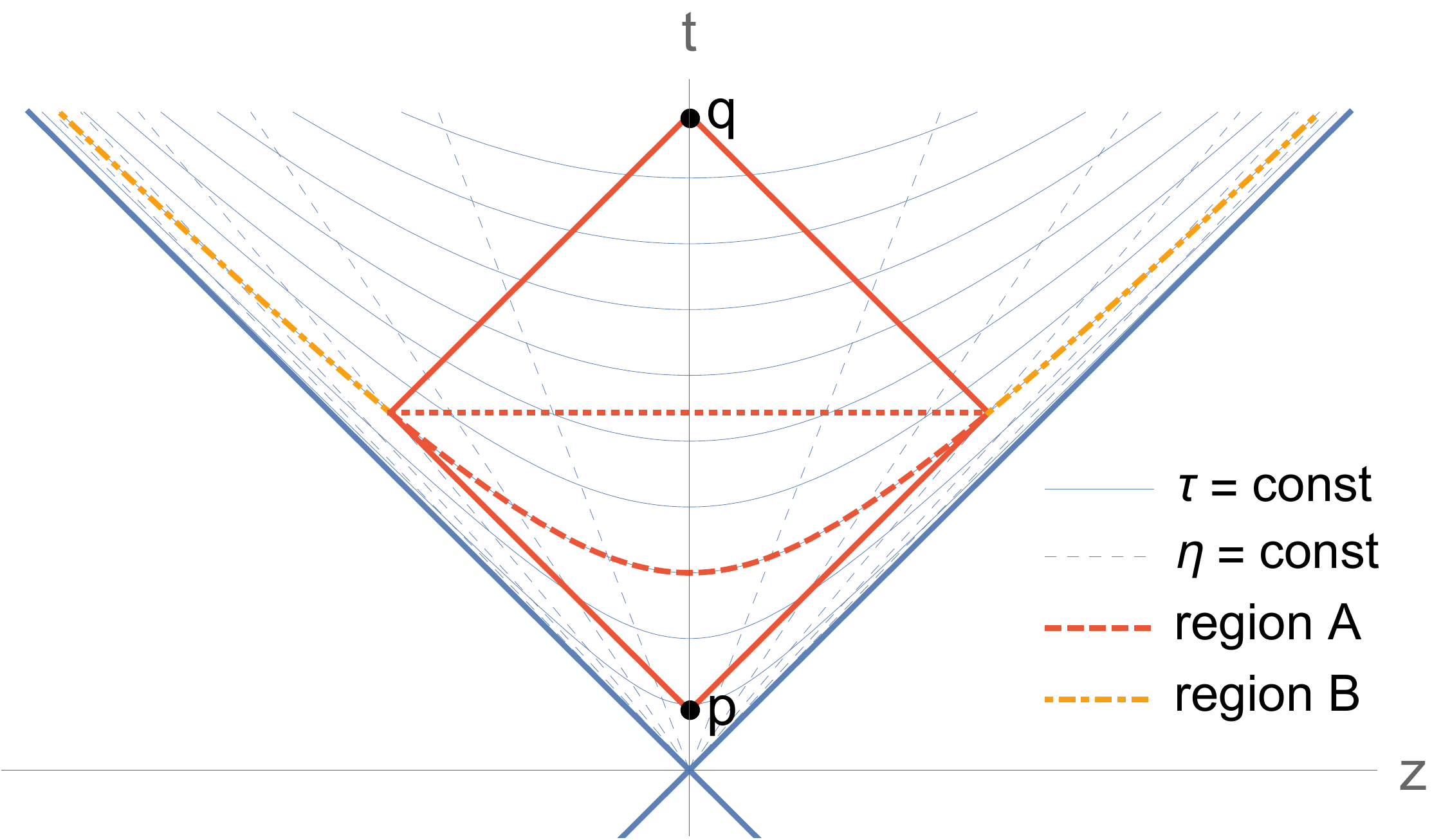}
	\caption{Illustration of Bjorken coordinates and causal development of a rapidity interval $(-\Delta\eta/2,\Delta\eta/2)$ at fixed proper time $\tau$ (region $A$, dashed red line). The complement region $B$ corresponds to $(-\infty,-\Delta\eta/2)$ and $(\Delta\eta/2, \infty)$ (dot-dashed orange line). The point $p$ is the origin of the past light cone that delimits region $A$ and $q$ is the endpoint of the future light cone. For better orientation we also show lines of constant proper time $\tau$ and rapidity $\eta$.}
	\label{fig:Geometry}
\end{figure}

The correlation functions on the finite rapidity interval can be most conveniently obtained from the momentum space representations \eqref{eq:426} or \eqref{eq:431}, together with the analog of the relation \eqref{eq:phinfromphip} expressing the discrete field basis in terms of standard Fourier modes with an appropriate integral kernel. In our case here, this becomes (at fixed time $\tau$)
\begin{equation}
\phi_n = \int \frac{dk}{2\pi} \sin(\tfrac{k \Delta\eta}{2}-\tfrac{n\pi}{2}) \left[ \frac{1}{k-\frac{n \pi}{\Delta\eta}} +  \frac{1}{k+\frac{n \pi}{\Delta\eta}} \right] \phi(k)\,.
\label{eq:kernel}
\end{equation} 

When one calculates correlators  in the discrete basis, such as  $\langle \phi_n \phi_{-m} \rangle_c$, from  \eqref{eq:phiphicorrelatorbarbasis} using the kernel \eqref{eq:kernel}, one observes that the terms proportional to the off-diagonal occupation number $\bar u(k)$ in \eqref{eq:phiphicorrelatorbarbasis} contain a term that oscillates very fast with $k$ in the limit $M \tau \to 0$. This is because the combination $2k \ln(\tau)+2 \theta(k,M)$, with the phase $\theta(k,M)$ in \eqref{eq:phaseTheta}, is strongly dependent on $k$ with infinite derivatives contributing in the limit $M\tau \to 0$. Hence these terms effectively do not contribute to the correlators such as $\langle \phi_n \phi_{-m} \rangle$ for a finite length interval $\Delta\eta$ in the limit $M\tau \to 0$.

One is then left with correlators that describe vacuum fluctuations and the occupation numbers $\bar n(k)$ corresponding to a thermal distribution at very early times $\tau$. This shows that the two limits $M\tau \to 0$ and $\Delta\eta \to \infty$ do not commute. If one considers $M\tau \to 0$ for finite $\Delta \eta$, one finds the entanglement entropy of a thermal state with the time dependent temperature given by \eqref{eq:Ttau}. This holds even if one then takes the limit $\Delta\eta\to \infty$. In contrast, if one considers an infinite interval $\Delta\eta \to \infty$ for finite $M\tau$, one finds a pure state with vanishing entanglement entropy. This holds also if one considers subsequently the limit of vanishing mass $M\tau\to 0$. 
\begin{figure}
	\centering
	\includegraphics[width=0.45\textwidth]{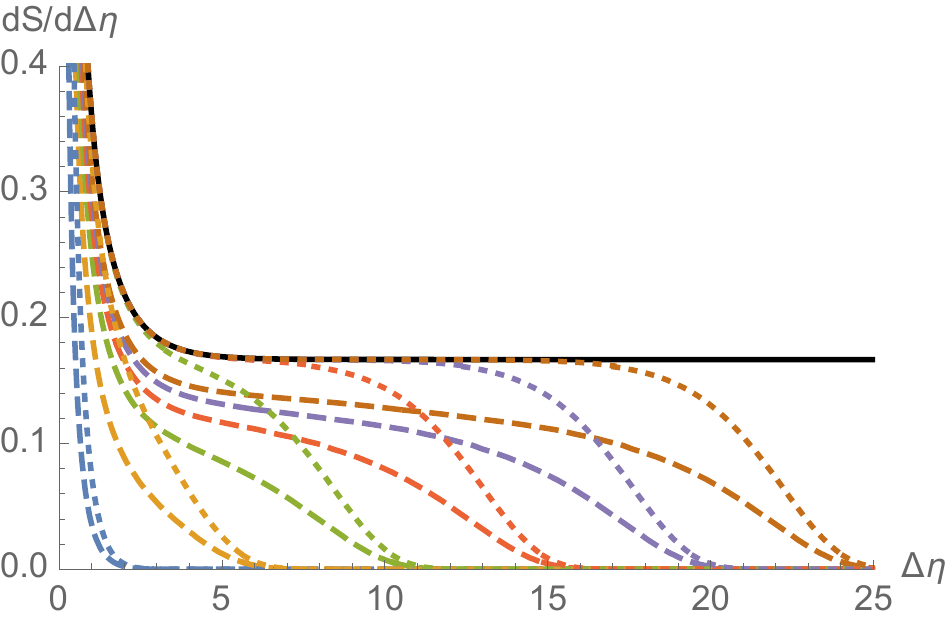}
	\caption{Entanglement entropy density as a function of the rapidity interval for the free massive scalar fields (dashed curves) and free massive Dirac fermions (dotted curves). From left to right, the curves correspond to $M\tau=1$, $M\tau=10^{-1}$, $M\tau=10^{-2}$, $M\tau=10^{-3}$, $M\tau=10^{-4}$, and $M\tau=10^{-5}$. At sufficiently early time, a plateau forms corresponding to the conformal case (solid black line).}
	\label{fig2}
\end{figure}

Remarkably, one finds that at the very early times $M\tau\to 0$, this entanglement entropy in a finite rapidity interval $\Delta\eta$ is equivalent to that of a 1+1 dimensional  conformal field theory at finite temperature when $T=1/(2\pi \tau)$. In such a 1+1 dimensional conformal field theory at temperature $T$, the entanglement entropy of an interval of length $L$ is given by \cite{Korepin:2004zz,Calabrese:2004eu}
\begin{equation}
S(T,l) = \frac{c}{3} \ln\left( \frac{1}{\pi T \epsilon} \sinh(\pi L T) \right) + \text{const}\,,
\end{equation}
with central charge $c=1$ in the present situation and where $\epsilon$ is a small length serving as ultraviolet regulator. Setting now $L=\tau \Delta \eta$ and $T=1/(2\pi \tau)$ leads to the result
\begin{equation}
S(\tau, \Delta\eta) = \frac{c}{3} \ln \left( 2\tau \,\text{sinh}(\Delta\eta/2)/\epsilon\right)+\text{const}.
\label{eq:EntEntConformalBjorken}
\end{equation}

Alternatively, one can obtain the result in \eqref{eq:EntEntConformalBjorken} for the expanding system from the Minkowski space  computations in our framework, as discussed in section \ref{sec:EntEntropyMinkowski}. In obtaining this result, we first make use of the fact that the coherent field does not contribute to the connected correlation functions in the covariance matrix and can be dropped from the computation of the entanglement entropy. Furthermore, the entanglement entropy of an interval is unchanged by unitary evolution as long as the boundaries are kept fixed. This allows one to reduce the problem to finding the entanglement entropy of an interval in Minkowski space at a constant time with length $\Delta z=L=2 \tau \sinh\left( \Delta\eta/2 \right)$ at a fixed time $t=\tau \cosh(\Delta\eta/2)$; this corresponds to the dotted red line in Fig.\ \ref{fig:Geometry}. Using \eqref{eq:LogLaw}, this leads to \eqref{eq:EntEntConformalBjorken} as well. This reasoning has been used for the derivation of the results first obtained in \cite{Berges:2017zws}.

The additive constant in eq.\ \eqref{eq:EntEntConformalBjorken} is not universal but the derivatives of $S$ with respect to $\tau$ and $\Delta\eta$ are universal. This implies $\tau \partial S/\partial\tau  = c/3$ and $\partial S/\partial\Delta\eta = (c/6)\, \text{coth}(\Delta\eta/2)$. For a large rapidity interval $\Delta\eta\gg 1$, one has $S = (c/6) [ \Delta\eta +2 \ln (\tau)]+\text{const}$.
This shows the existence of  a time-independent piece of the entanglement entropy that is extensive in rapidity and a  $\Delta\eta$-independent piece that grows logarithmically with the proper time. 

At later times, for the nonconformal case of free massive scalars, the universal part of the entanglement entropy behaves as in the conformal case for $M\Delta z \ll 1$ and decays for $M \Delta z \gg 1$ \cite{Casini:2009sr}. In this case, the derivative with respect to $\Delta\eta$ of the entanglement entropy gives
\begin{equation}
\begin{split}
& \frac{\partial}{\partial \Delta \eta}S(\tau, \Delta\eta) =  \frac{\partial S}{\partial \ln \Delta z} \frac{\partial \ln \Delta z}{\partial \Delta\eta} \\
& =  c_E\left(2M\tau \sinh(\Delta\eta/2)\right)  \coth(\Delta\eta/2)/2\,,
\end{split}
\end{equation}
where $c_E\left(M\Delta z\right) = \Delta z {\partial S}/{\partial \Delta z}$ is the entanglement entropy $c$-function for a massive scalar field. Taking the  conformal limit, one obtains, as anticipated, that $c_E(0)=c/3=1/3$. For large values of the argument, this function has the form $c_E(x)\to x K_1(2x)/4$, which decays exponentially. 

We can employ the general expression for $c_E(x)$~\cite{Casini:2009sr} for massive free bosons to compute $dS/d\Delta\eta$ for the massless Schwinger model. The result is displayed in Fig.\ \ref{fig2} for the different  values of $M\tau$ that are shown in the caption (dashed lines). For short times $\tau M \ll 1$, one observes a significant entanglement over rapidity intervals $\Delta\eta={\cal O}(1)$. At intermediate values of $\tau M$ and $\Delta\eta$, one observes that $\partial S/\partial \Delta\eta$ approaches a plateau at $1/6$ as a function of $\Delta\eta$ at early times. One also sees that it decays  both for very large $\Delta\eta$ and for later times $\tau$. The plateau is governed by the conformal limit $M\tau\rightarrow 0$ which is shown by the solid black line in the figure. 

Since the entanglement entropy $c$-function in the conformal limit is identical for real massless scalar bosons and for massless Dirac fermions, this indicates that the conformal limit is consistently described in the Schwinger model with or without bosonization. One may also determine the entanglement entropy of free massive Dirac fermions using similar manipulations as described above and using the corresponding $c$-function given in Ref.~\cite{Casini:2009sr}. The result is also shown in Fig.\ \ref{fig2} (dotted lines). The approach to the universal plateau at $dS/d\Delta\eta=1/6$ for $M\tau\to 0$ is even faster for the free fermion case.

A more intuitive (but also more heuristic) description of why the vacuum state $|\Omega\rangle$ in the conformal limit looks thermal in a finite rapidity interval $\Delta\eta$ is as follows. Although a pure state, the state describing the expanding string contains entangled pairs of quasiparticles with opposite momentum in the Bogoliubov basis \eqref{eq:BesselModeFunctions} which consists of modes with positive frequencies with respect to $\ln(\tau)$ for $M\to 0$. This particular basis is special because only there can one interpret quasiparticles (in the classical quasiparticle limit) as moving in space on well defined trajectories. For a rapidity interval $\Delta\eta$, this implies that quasiparticles constantly come in via the left and right boundaries. They are entangled with other quasiparticles moving in the opposite direction but that is not seen locally. Because these quasiparticles have a thermal spectrum, local observables will effectively look thermal. A related argument was employed previously to understand the time evolution of entanglement entropy after a quantum quench \cite{Calabrese:2005in, Calabrese:2016xau}.

\subsection{Local density matrix of an expanding string}
\label{sec:LocalDensityMatrix}

One can go even further and make stronger statements regarding the thermal character of entanglement entropy in a finite interval of the expanding string. Note that the correlators in \eqref{eq:phiphicorrelatorbarbasis}, when projected to the finite interval with the kernel \eqref{eq:kernel}, are at $M\tau\to 0$ {\it exactly} those of the 1+1-dimensional conformal field theory in thermal equilibrium if the temperature is identified to be $T=1/(2\pi \tau)$. Because Gaussian density matrices are fully specified by one-point expectation values and two-point correlation functions, the density matrix obtained by first taking the limit $M\tau \to 0$ and then $\Delta\eta\to \infty$ has the thermal form,
\begin{equation}
\rho = \frac{1}{Z} e^{-K}, \quad\quad\quad Z = \text{Tr} \; e^{-K},
\label{eq:rhointermsofK}
\end{equation}
where $K$ is the so-called modular or entanglement Hamiltonian defined as 
\begin{equation}
K = \int_\Sigma d\Sigma^\mu \beta^\nu T_{\mu\nu} \, .
\label{eq:LocalModularHamiltonian}
\end{equation}
Equations \eqref{eq:rhointermsofK} and \eqref{eq:LocalModularHamiltonian} specify the density matrix of a locally thermal state where $\beta^\mu = u^\mu / T$ is the so-called inverse temperature vector given by the fluid velocity $u^\mu$ (pointing in $\tau$-direction) and the temperature is $T=1/(2\pi \tau)$. Moreover, $T_{\mu\nu}$ is the energy-momentum tensor of excitations $\varphi(\tau, \eta)$ around the coherent field $\bar \phi(\tau)$ according to \eqref{eq:fieldsplitting}. Note also that $\beta^\mu=(1/T, 0)$ (in coordinates $\tau, \eta$) is actually a conformal Killing vector according to \eqref{eq:conformalKillingEqn}. 

This result can also be obtained as a limit of a more general result, as discussed already in Ref.\ \cite{Berges:2017zws}. A conformal field theory in the vacuum state in a region with a boundary formed by the intersection of two light cones (see figure \ref{fig:Geometry}), has a reduced density matrix of the form \eqref{eq:rhointermsofK} on any hypersurface in that region that has the same boundary. The modular Hamiltonian is a local expression given by~\cite{Casini:2011kv,Arias:2016nip} (see also \cite{Candelas:1978gf})
\begin{equation}
K = \int_\Sigma d\Sigma^\mu \xi^\nu T_{\mu\nu} \, .
\label{eq:LocalModularHamiltonian}
\end{equation}
Here $T_{\mu\nu}$ is again the energy-momentum tensor of excitations and $\xi^\nu$ is a vector field that can be written as 
\begin{equation}
\xi^\mu = \frac{2\pi}{(q-p)^2} \left[ (q-x)^\mu (x-p)(q-p) + (x-p)^\mu (q-x)(q-p) - (q-p)^\mu (x-p)(q-x) \right]\,,
\end{equation}
where $x$ is the space-time position on the hypersuface we consider, $q$ is the end-point of the future light cone and $p$ the starting point of the past light cone that form the boundary of the region.
Note that \eqref{eq:LocalModularHamiltonian} is again of the same form as a density matrix of a thermal state if one identifies $ \xi^\mu = \frac{1}{T}u^\mu $ the vector of (inverse of) temperature $T$ and fluid velocity $u^\mu$. The vector $\xi^\mu$ vanishes on the boundary of the region enclosed by the two light cones corresponding formally to an infinite temperature. 

Consider now a situation where the two enclosing light cones intersect on a constant $\tau$ hypersurface in Bjorken coordinates with a rapidity difference $\Delta \eta$. If we concentrate on the point in the middle of this rapidity interval, the vector $\xi^\mu$ points in $\tau$-direction and has length $2\pi \tau (\cosh(\Delta\eta/2)-1)/\sinh(\Delta\eta/2)$. The associated temperature approaches precisely $T=1/(2\pi \tau)$ for $\Delta\eta \to \infty$.

The state $\rho$ we are considering here is vacuum-like but may have a nonvanishing coherent field and therefore, a corresponding nonzero energy. Further, excitations that can be formed from local unitary (entropy preserving) transformations lead to a modified density matrix $\rho^1$. Using the notion of a relative entropy  to $\rho$ as explained in appendix \ref{app:relativeentropy}, we have
\begin{equation}
S(\rho^1 | \rho) = \Delta \langle K \rangle = \beta^\mu P_\mu \, ,
\end{equation}
where $P_\mu=\int_\Sigma d\Sigma^\nu \langle T_{\mu\nu} \rangle$ is the four-momentum associated with the perturbation. This is precisely the characteristic of a thermal state. 

Within a large but finite rapidity interval, the relative entropy of the state with a small perturbation compared to the coherent field state of the expanding string has the same value as in a thermal state with temperature $T=1/(2\pi \tau)$. In the expanding geometry of interest, quantum fluctuations of this kind are therefore as likely as in a thermal state. Such fluctuations should therefore be observed with a distribution corresponding to that of a grand canonical ensemble.

\section{Conclusions}
\label{sec:conclusions}

We developed in this paper a powerful formalism exploiting Gaussian density matrices to examine different entanglement entropies that arise in a wide variety of equilibrium and nonequilibrium problems in quantum field theory. The most important results of this exercise are expressions for the entanglement entropy in terms of two-point correlation functions that can be directly evaluated within the finite spacetime interval of interest. 

In particular, we revisited the computation of the entanglement entropy of an interval of length $L$ in static, two-dimensional Minkowski space and discussed how it can be computed within our formalism. We find that it can be fully determined by correlation functions within the interval. Our results clearly reveal the importance of boundary conditions in the emergence of the entanglement entropy in finite spacetime intervals.

We then applied the insights gain from our general treatment to study entanglement in the expanding string formed between a highly energetic quark-antiquark pair after an electron-positron collision. In order to do so, we employed the Schwinger model of quantum electrodynamics, which is the basis of much of the phenomenology describing multiparticle production in these collisions. We exploit the fact that the Schwinger model can be bosonized and is (in the limit of vanishing fermion mass $m$) equivalent to a free bosonic theory. The expanding string is best described in Bjorken coordinates of proper time $\tau$ and rapidity $\eta$ and corresponds then to a coherent field solution of the Klein-Gordon equation in an expanding geometry. The entanglement properties of this state are fixed in terms of correlation functions for excitations around the coherent field. We discuss them in terms of appropriate mode functions. We observe that different sets of mode functions are possible and that they are related by Bogoliubov transformations. 

We find that the state corresponding to the standard Minkowski space vacuum appears at early time $\tau$ as an occupied squeezed state with mode functions that have positive frequency with respect to the logarithm of Bjorken time $\ln(\tau)$. Moreover, as we showed very explicitly in sections \ref{sec:DynamicsPerturbations} - \ref{sec:LocalDensityMatrix} and had discussed previously in \cite{Berges:2017zws}, this state appears in any finite rapidity interval as a thermal state governed by a $\tau$-dependent temperature $T=\hbar / (2\pi\tau)$. This is a rather remarkable finding, since it implies that excitations around the coherent field solution of the expanding sting are in fact thermal at very early times! Although our model for a QCD string is not fully realistic, we believe that the above described mechanism provides a compelling candidate for a deeper understanding of the approximately thermal distributions of hadron ratios found in many collider experiments even in contexts where scattering effects contributing to thermalization of particles in the final state are likely small. 

A crucial question beyond the framework investigated here is what happens when interactions are taken into account. For the bosonized massless Schwinger model, they arise  from a nonvanishing fermion mass $m$, leading to a Sine-Gordon type interaction term as in eq.\ \eqref{eq:SchwingerModelBosonizedText}. At very early Bjorken time $\tau$, the interaction term $\sim m e$ is expected to be irrelevant but could start to play a role at intermediate times and -- for large enough $m$ -- even before the boson mass term $\sim e^2/\pi$. We expect that further correlations, and therefore entanglement,  can be built up by interactions between excitations and we plan to investigate their consequences for the entanglement dynamics in future work. 

While we have concentrated here on dynamics in 1+1 dimensions, similar entanglement dynamics may be at play in a Bjorken-type expanding geometry with additional transverse dimensions. If this is the case, quantum entanglement may be an important feature of early time dynamics in heavy-ion collisions. We plan to explore this exciting possibility in future work.

Similar considerations also apply to nonrelativistic condensed matter systems such as ultra-cold quantum gases whose ``horizon'' is set by the speed of sound. Entanglement in such many-body systems can be efficiently explored with table-top experiments enabling direct confrontation of theoretical predictions with experimental measurements.
\begin{appendix}
\section{Relative entropy}
\label{app:relativeentropy}

Extending the discussion of section~\ref{sec:Entropies}, we now consider {\it two} density matrices $\rho$ and $\sigma$. They may be reduced density matrices originating from the trace over some part of the Hilbert space. We will be interested in the quantum relative entropy or Kullback-Leibler divergence of $\rho$ with respect to $\sigma$, defined by
\begin{equation}
S(\rho | \sigma) = \text{Tr} \{ \rho \, (\text{ln}\; \rho - \text{ln} \; \sigma) \}.
\label{eq:DefRelEntropy}
\end{equation}
The relative entropy measures in an information theoretic sense to which extend one can distinguish the distribution $\rho$ from the distribution $\sigma$, see \cite{Vedral:2002zz} for a review. In particular, for $S(\rho|\sigma)=0$, the two distributions cannot be distinguished and agree, $\rho=\sigma$ (up to a set of measure zero). For any other choice $\rho\neq\sigma$, the relative entropy is positive, $S(\rho| \sigma)>0$. 

An interesting special case is when $\sigma$ is the thermal density matrix
\begin{equation}
\sigma= \frac{1}{Z} e^{-\beta H},
\end{equation}
with $\beta=1/T$ and Hamiltonian $H$. The partition function can be written as $Z=e^{-\beta F}$ with free energy $F=E-TS$, $dF=-SdT-pdV$. The relative entropy becomes
\begin{equation}
\begin{split}
S(\rho | \sigma) = & \text{Tr} \rho \; \text{ln} \; \rho + \beta \left( \langle H \rangle_\rho - F_\sigma \right) \\
= & - S_\rho + S_\sigma + \beta \left( \langle  H \rangle_\rho - \langle H \rangle_\sigma \right),
\end{split}
\label{eq:SrhosigmaThermal}
\end{equation}
where $\langle \cdot \rangle_\rho$ and $\langle \cdot \rangle_\sigma$ denote expectation values with respect to the density matrices $\rho$ and $\sigma$, respectively, while $S_\rho$ and $S_\sigma$ denote the corresponding von Neumann entropies.
Possible UV divergent contributions to the entanglement entropy are independent of the state and cancel between the first and second term in the second line of \eqref{eq:SrhosigmaThermal}. (This is actually a general statement independent of the specific choice for $\sigma$ made here.) Moreover, also possibly UV divergent contributions to the expectation values of energy, e.\ g.\ from the zero-point fluctuations of various modes cancel on the right hand side of \eqref{eq:SrhosigmaThermal}. Note that for equal energy, $\langle  H \rangle_\rho = \langle H \rangle_\sigma$, the relative entropy \eqref{eq:SrhosigmaThermal} equals the the difference of entropies.

One may also consider a situation where $\rho$ can be obtained from $\sigma$ by a unitary operation. In that case $S_\rho=S_\sigma$ and $S(\rho | \sigma)=\beta (\langle H \rangle_\rho - \langle H \rangle_\sigma)$, which is reminiscent of the definition of temperature starting from the microcanonical ensemble, $dS=\beta dE$. These considerations can be extended away from the case where $\sigma$ is thermal in terms of the modular Hamiltonian $K=-\ln \sigma$, $S(\rho|\sigma)=\langle K \rangle_\rho - \langle K \rangle_\sigma\geq 0$.

Similar as for the von Neumann entropy we will perform the calculation in the replica formalism and obtain the relative entropy in \eqref{eq:DefRelEntropy} as a limit of the R\'{e}nyi relative entropies \cite{Lashkari:2014yva} (the latter defined in \cite{Ruggiero:2016khg})
\begin{equation}
S_N(\rho | \sigma) = \frac{1}{1-N} \ln \frac{\text{Tr}\{ \rho\, \sigma^{N-1} \}}{\text{Tr}\{ \rho^N \}} = \frac{1}{1-N} \ln \text{Tr}\{ \rho \, \sigma^{N-1} \}- S_N(\rho).
\end{equation}
With these definitions one has \cite{Lashkari:2014yva}
\begin{equation}
S(\rho | \sigma) = \lim_{N\to 1} S_N(\rho | \sigma).
\end{equation}
When both $\rho$ and $\sigma$ are Gaussian density matrices, they are fully characterized by their respective expectation values
\begin{equation}
\begin{split}
\langle \phi_m \rangle_\rho = & \text{Tr}\{ \rho \, \phi_m \} = \bar \phi_m^{(\rho)},  \quad\quad\quad
\langle \pi_m \rangle_\rho = \text{Tr}\{ \rho \, \pi_m \} = j_m^{(\rho)}, \\
\langle \phi_m \rangle_\sigma = & \text{Tr}\{ \sigma \, \phi_m \} = \bar \phi_m^{(\sigma)},  \quad\quad\quad
\langle \pi_m \rangle_\sigma = \text{Tr}\{ \sigma \, \pi_m \} = j_m^{(\sigma)},
\end{split}
\end{equation}
as well as connected two-field correlation functions, specified for $\rho$ in terms of $h^{(\rho)}$, $h^{(\rho)\dagger}$, $\Sigma_a^{(\rho)}$ and $\Sigma_b^{(\rho)}$ and similar for $\sigma$ in terms of $h^{(\sigma)}$, $h^{(\sigma)\dagger}$, $\Sigma_a^{(\sigma)}$ and $\Sigma_b^{(\sigma)}$ (see discussion in section \ref{sec:Corrfuncts}).

For simplicity we will assume in the following that the field expectation values with respect to $\rho$ and $\sigma$ agree, i.\ e.\ $\bar\phi_m^{(\rho)} = \bar\phi_m^{(\sigma)}$ and $j_m^{(\rho)} = j_m^{(\sigma)}$. We make no such assumption about the connected correlation functions, however, and in fact keep them fully general. 

Using similar manipulations as in section \ref{sec:Entropies} we obtain
\begin{equation}
\begin{split}
 \frac{1}{1-N} & \text{ln} \; \text{Tr} \{ \rho \, \sigma^{N-1} \} = \\
&  \frac{1}{2(N-1)} {\Bigg \{} 
 \text{Tr} \ln  {\Bigg (} 
 {\bigg(} \sqrt{\tfrac{1}{4}+a^{(\rho)}+b^{(\rho)2}}+b^{(\rho)}+\frac{1}{2} {\bigg )}
 {\bigg(} \sqrt{\tfrac{1}{4}+a^{(\sigma)}+b^{(\sigma)2}}+b^{(\sigma)}+\frac{1}{2} {\bigg )}^{N-1} \\
 & \quad\quad\quad\quad\quad\quad\quad - 
  {\bigg(} \sqrt{\tfrac{1}{4}+a^{(\rho)}+b^{(\rho)2}}+b^{(\rho)}-\frac{1}{2} {\bigg )}
 {\bigg(} \sqrt{\tfrac{1}{4}+a^{(\sigma)}+b^{(\sigma)2}}+b^{(\sigma)}-\frac{1}{2} {\bigg )}^{N-1} {\Bigg )} \\
&  \quad\quad\quad\quad +\text{Tr} \ln  {\Bigg (} 
 {\bigg(} \sqrt{\tfrac{1}{4}+a^{(\rho)}+b^{(\rho)2}}-b^{(\rho)}+\frac{1}{2} {\bigg )}
 {\bigg(} \sqrt{\tfrac{1}{4}+a^{(\sigma)}+b^{(\sigma)2}}-b^{(\sigma)}+\frac{1}{2} {\bigg )}^{N-1} \\
 & \quad\quad\quad\quad\quad\quad\quad - 
  {\bigg(} \sqrt{\tfrac{1}{4}+a^{(\rho)}+b^{(\rho)2}}-b^{(\rho)}-\frac{1}{2} {\bigg )}
 {\bigg(} \sqrt{\tfrac{1}{4}+a^{(\sigma)}+b^{(\sigma)2}}-b^{(\sigma)}-\frac{1}{2} {\bigg )}^{N-1} 
 {\Bigg )} {\Bigg \} },
\end{split}
\label{eq:relEntropSecPart}
\end{equation}
with
\begin{equation}
\begin{split}
a^{(\rho)} = & {\big (}h^{(\rho)} + h^{(\rho)\dagger} - 2 \Sigma_a^{(\rho)}{\big )}^{-1} \Sigma_a^{(\rho)}, \\
b^{(\rho)} = & {\big (}h^{(\rho)} + h^{(\rho)\dagger} - 2 \Sigma_a^{(\rho)}{\big )}^{-1} \Sigma_b^{(\rho)}, \\
a^{(\sigma)} = & {\big (}h^{(\sigma)} + h^{(\sigma)\dagger} - 2 \Sigma_a^{(\sigma)}{\big )}^{-1} \Sigma_a^{(\sigma)}, \\
b^{(\sigma)} = & {\big (}h^{(\sigma)} + h^{(\sigma)\dagger} - 2 \Sigma_a^{(\sigma)}{\big )}^{-1} \Sigma_b^{(\sigma)}.
\end{split}
\end{equation}
By subtracting \eqref{eq:RenyiEntropy} from \eqref{eq:relEntropSecPart} one obtains directly an expression for the relative R\'{e}nyi entropy. In particular, we obtain for the quantum relative entropy by taking the limit $N\to 1$, the expression
\begin{equation}
\begin{split}
S(\rho | \sigma) = 
  \frac{1}{2}\text{Tr} & {\Bigg \{}  \left(\sqrt{\tfrac{1}{4}+a^{(\rho)}+b^{(\rho)2}}+b^{(\rho)}+\tfrac{1}{2} \right) {\Big [} \ln {\Big (}\sqrt{\tfrac{1}{4}+a^{(\sigma)}+b^{(\sigma)2}}+b^{(\sigma)}+\tfrac{1}{2} {\Big )}  \\
 & \quad\quad\quad\quad\quad\quad\quad\quad\quad\quad\quad\quad\quad\quad\quad - \ln {\Big (}\sqrt{\tfrac{1}{4}+a^{(\rho)}+b^{(\rho)2}}+b^{(\rho)}+\tfrac{1}{2} {\Big )} {\Big ]}\\
& - \left(\sqrt{\tfrac{1}{4}+a^{(\rho)}+b^{(\rho)2}}+b^{(\rho)}-\tfrac{1}{2} \right) {\Big [} \ln {\Big (}\sqrt{\tfrac{1}{4}+a^{(\sigma)}+b^{(\sigma)2}}+b^{(\sigma)}-\tfrac{1}{2} {\Big )}  \\
 & \quad\quad\quad\quad\quad\quad\quad\quad\quad\quad\quad\quad\quad\quad\quad - \ln {\Big (}\sqrt{\tfrac{1}{4}+a^{(\rho)}+b^{(\rho)2}}+b^{(\rho)}-\tfrac{1}{2} {\Big )} {\Big ]}\\
& +\left(\sqrt{\tfrac{1}{4}+a^{(\rho)}+b^{(\rho)2}}-b^{(\rho)}+\tfrac{1}{2} \right) {\Big [} \ln {\Big (}\sqrt{\tfrac{1}{4}+a^{(\sigma)}+b^{(\sigma)2}}-b^{(\sigma)}+\tfrac{1}{2} {\Big )}  \\
 & \quad\quad\quad\quad\quad\quad\quad\quad\quad\quad\quad\quad\quad\quad\quad - \ln {\Big (}\sqrt{\tfrac{1}{4}+a^{(\rho)}+b^{(\rho)2}}-b^{(\rho)}+\tfrac{1}{2} {\Big )} {\Big ]}\\
& - \left(\sqrt{\tfrac{1}{4}+a^{(\rho)}+b^{(\rho)2}}-b^{(\rho)}-\tfrac{1}{2} \right) {\Big [} \ln {\Big (}\sqrt{\tfrac{1}{4}+a^{(\sigma)}+b^{(\sigma)2}}-b^{(\sigma)}-\tfrac{1}{2} {\Big )}  \\
 & \quad\quad\quad\quad\quad\quad\quad\quad\quad\quad\quad\quad\quad\quad\quad - \ln {\Big (}\sqrt{\tfrac{1}{4}+a^{(\rho)}+b^{(\rho)2}}-b^{(\rho)}-\tfrac{1}{2} {\Big )} {\Big ]} {\Bigg \}}.
\end{split}
\end{equation}
In the special case of $b^{(\sigma)}=b^{(\rho)}=0$ this simplifies to
\begin{equation}
\begin{split}
S(\rho| \sigma) = \text{Tr} & {\Bigg \{}  \left(\sqrt{\tfrac{1}{4}+a^{(\rho)}}+\tfrac{1}{2} \right) {\Big [} \ln {\Big (}\sqrt{\tfrac{1}{4}+a^{(\sigma)}}+\tfrac{1}{2} {\Big )}  - \ln {\Big (}\sqrt{\tfrac{1}{4}+a^{(\rho)}}+\tfrac{1}{2} {\Big )} {\Big ]}\\
& - \left(\sqrt{\tfrac{1}{4}+a^{(\rho)}}-\tfrac{1}{2} \right) {\Big [} \ln {\Big (}\sqrt{\tfrac{1}{4}+a^{(\sigma)}}-\tfrac{1}{2} {\Big )} - \ln {\Big (}\sqrt{\tfrac{1}{4}+a^{(\rho)}}-\tfrac{1}{2} {\Big )} {\Big ]} {\Bigg \}}.
\end{split}
\end{equation}

\section{Symmetries, anomalies and bosonization}
\label{app:A}

In this appendix we recall how the Schwinger model corresponding to vector-like QED in two dimensions can be bosonized \cite{Coleman:1975pw}. One proceeds via a discussion of gauge symmetries and associated quantum anomalies. 

\subsection{Symmetries, conservation laws and anomalies}
Consider the Lagrangian
\begin{equation}
\mathscr{L} = - \bar \psi ( \gamma^\mu \partial_\mu - i e A_\mu \gamma^\mu - i e B_\mu \gamma^\mu \gamma_5 ) \psi.
\end{equation}
The vector gauge symmetry is as usual
\begin{equation}
\psi \to e^{i\alpha} \psi, \quad\quad \bar \psi \to \bar \psi e^{-i\alpha}, \quad\quad A_\mu \to A_\mu + \frac{1}{e} \partial_\mu \alpha, \quad\quad B_\mu \to B_\mu,
\end{equation}
or for left and right handed fermions
\begin{equation}
\psi_{L} \to e^{i\alpha} \psi_{L}, \quad\quad \psi_{R} \to e^{i\alpha} \psi_{R}, \quad\quad  \psi_{L}^* \to \psi_{L}^* e^{-i\alpha}, \quad\quad \psi_{R}^* \to \psi_{R}^* e^{-i\alpha} .
\end{equation}
One can define the regularization such that the vector gauge symmetry above is not anomalous. The associated Noether vector current is in our notation
\begin{equation}
J^\mu = i \bar \psi \gamma^\mu \psi,
\end{equation}
and it is classically and quantum theoretically conserved, $\partial_\mu \langle J^\mu \rangle=0$.

In contrast, the axial gauge transformation
\begin{equation}
\psi \to e^{i\beta \gamma_5} \psi, \quad\quad \bar \psi \to \bar \psi e^{i\beta \gamma_5}, \quad\quad A_\mu \to A_\mu, \quad\quad B_\mu \to B_\mu + \frac{1}{e} \partial_\mu \beta ,
\end{equation}
or for left and right handed fermions
\begin{equation}
\psi_{L} \to e^{i\beta} \psi_{L}, \quad\quad \psi_{R} \to e^{-i\beta} \psi_{R},  \quad\quad  \psi_{L}^* \to \psi_{L}^* e^{-i\beta}, \quad\quad  \psi_{R}^* \to \psi_{R}^* e^{i\beta},
\end{equation}
is anomalous. The axial vector current
\begin{equation}
J_5^\mu = i \bar \psi \gamma^\mu \gamma_5 \psi
\end{equation}
is not conserved. There is an associated anomaly (see e.\ g.\ ref.\ \cite{Anomalies}, eq.\ (5.41))
\begin{equation}
\partial_\mu \langle J^\mu_5 \rangle = - \frac{1}{2\pi} \epsilon^{\mu\nu} e F_{\mu\nu} - \frac{1}{\pi} e \partial_\mu B^\mu .
\end{equation}
The Schwinger functional defined by
\begin{equation}
e^{iW}= \int D\bar\psi D \psi \; e^{i \int d^2 x \mathscr{L}}
\end{equation}
changes by an infinitesimal axial transformation $\delta \beta$ according to
\begin{equation}
\delta W= - \int d^2 x \, \delta\beta(x) \partial_\mu \langle j_5^\mu(x) \rangle =  \int d^2x \left\{ \delta\beta(x) \frac{e}{2\pi}\epsilon^{\mu\nu} F_{\mu\nu}(x) + \delta \beta \frac{e}{\pi} \partial_\mu B^\mu \right\}.
\end{equation}
\subsection{Bosonization}
Consider now the partition function
\begin{equation}
e^{iW[J]} = \int D\phi D\bar \psi D \psi D A \; e^{iS+i\int_x J^\mu A_\mu }
\end{equation}
where $S=\int_x \mathscr{L}$ is essentially the microscopic action of the mass-less Schwinger model and we have added an irrelevant constant in the form of an integral over a free bosonic field $\phi$. Decompose the gauge field as
\begin{equation}
A_\mu = \frac{1}{e}\partial_\mu \zeta + \frac{1}{e}\partial^\nu \xi \, \epsilon_{\nu\mu},
\end{equation}
with two functions $\zeta$ and $\xi$. The Lagrangian is
\begin{equation}
\mathscr{L} = -\frac{1}{2}\partial_\mu\phi\partial^\mu\phi - \bar \psi \left(\gamma^\mu \partial_\mu - i \partial_\mu \zeta \gamma^\mu - i \partial_\mu \xi \gamma^\mu \gamma_5 \right) \psi - \tfrac{1}{4}F_{\mu\nu} F^{\mu\nu}.
\end{equation}
The idea is now to perform now a vector gauge transformation with $\alpha= -\zeta$ (which is simple to do) and an axial vector gauge transformation with $\beta = -\xi$. One needs to do this in infinitesimal steps taking into account the presence of the axial vector potential $B_\mu=\partial_\mu (\xi+\beta)$. 
The integration of the anomaly gives 
\begin{equation}
\int_0^{-\xi} d\beta \left\{ \frac{e}{2\pi} \epsilon^{\mu\nu} F_{\mu\nu} + \frac{e}{\pi} \partial_\mu\partial^\mu (\xi+\beta) \right\} = - \frac{e}{2\pi} \xi \epsilon^{\mu\nu}F_{\mu\nu} - \frac{e}{2\pi} \xi \partial_\mu \partial^\mu \xi.
\end{equation}
The fermionic functional integral becomes now a free one which contributes only an irrelevant factor to the partition function. Taking the anomaly into account leads therefore to
\begin{equation}
e^{iW[J]} = \int D\phi D A \, e^{i \int_x \left\{ - \frac{1}{2}\partial_\mu \phi \partial^\mu\phi -\frac{1}{4}F_{\mu\nu}F^{\mu\nu} - \frac{e}{2\pi} \xi \epsilon^{\mu\nu} F_{\mu\nu} - \frac{e}{2\pi} \xi \partial_\mu \partial^\mu \xi + J^\mu A_\mu\right\}} .
\end{equation}
Use now $A_\mu=\frac{1}{e}\partial^\nu \xi \epsilon_{\nu\mu}$ which implies $\epsilon^{\mu\nu}e F_{\mu\nu}= -2 \partial_\mu\partial^\mu \xi$, perform a shift of variables $\phi+\frac{1}{\sqrt{\pi}} \xi \to \phi$ and use $\partial^\mu \xi = -e \epsilon^{\mu\nu} A_\nu$ to obtain
\begin{equation}
e^{iW[J]} = \int D\phi DA \, e^{i \int_x \left\{- \frac{1}{2}\partial_\mu \phi \partial^\mu \phi -\frac{1}{4}F_{\mu\nu} F^{\mu\nu} -\frac{e}{\sqrt{\pi}}\partial_\mu \phi \, \epsilon^{\mu\nu}A_\nu+ J^\mu A_\mu\right\}}.
\end{equation}
This constitutes the bosonized form of the Schwinger model. In last step one can integrate out the gauge field by performing the Gaussian integral or, equivalently, solving the corresponding field equation.

In two dimensions, the gauge field $A_\mu$ is not dynamical. One may chose the axial gauge $A_1=0$ and the field strength is $F_{10}=-F_{01}=E_1=\partial_1 A_0$. (There is no magnetic field in two dimensions.)
The equation of motion is in the bosonized theory
\begin{equation}
\partial_\mu F^{\mu\nu} 
= \frac{e}{\sqrt{\pi}} \partial_\mu \phi \epsilon^{\mu\nu},
\end{equation}
and one can formally solve this
\begin{equation}
F^{\mu\nu} = e \left( \frac{1}{\sqrt{\pi}} \phi + \frac{\theta}{2\pi} \right) \epsilon^{\mu\nu},
\end{equation}
where $\theta$ appears as an integration constant and is the vacuum angle. Using this in the partition function gives (for $J=0$)
\begin{equation}
e^{iW} = \int D\phi \, e^{i \int_x \left\{- \frac{1}{2}\partial_\mu \phi \partial^\mu \phi - \frac{1}{2} \frac{e^2}{\pi}\phi^2 + \frac{e^2}{8\pi^2}\theta^2\right\}}.
\end{equation}
This is now the partition function of a massive scalar particle with mass $m=\frac{e}{\sqrt{\pi}}$. The vacuum angle is here just an irrelevant overall factor and therefore drops out.

For the massive Schwinger model the standard bosonization is more involved because the mass term transforms non-trivially under the axial gauge transformations. Order by order in a perturbative series in $m$ one finds that the theory is equivalent to the bosonic theory \cite{Coleman:1975pw},
\begin{equation}
\mathscr{L} = - \frac{1}{2}\partial_\mu \phi \partial^\mu \phi - \frac{1}{2}\frac{e^2}{\pi} \phi^2 - \frac{m e \exp(\gamma)}{2\pi^{3/2}} \cos\left(2\sqrt{\pi}\phi + \theta\right).
\label{eq:MassiveSchwingerModelBosonizedLagrangian}
\end{equation} 
Here, $\gamma$ is the Euler constant but note that the factor in front of the cosine term could be absorbed into a rescaling of the fermion mass $m$. This is now an interacting theory, and it depends on the vacuum angle $\theta$.

\end{appendix}
\acknowledgments
This work is part of and supported by the DFG Collaborative Research Centre ``SFB 1225 (ISOQUANT)''. R.~V.'s research is supported by the U.\ S.\ Department of Energy Office of Science, Office of Nuclear Physics, under contracts No.\ DE-SC0012704. R.~V.\ would like to thank ITP Heidelberg and the Alexander von Humboldt Foundation for support, and ITP Heidelberg for their kind hospitality.

\end{document}